\documentclass{article}
\usepackage[utf8]{inputenc}
\usepackage[a4paper, total={6in, 8in}]{geometry}
\usepackage{xcolor}
\usepackage{amsmath}
\usepackage{amsfonts}
\usepackage{mathrsfs}
\usepackage{graphicx}
\usepackage{authblk}
\usepackage{tikz}
\usepackage{physics}
\usepackage{hyperref}
\usepackage{cite}
\usepackage{simpler-wick}
\usepackage[export]{adjustbox}
\usetikzlibrary{matrix,arrows,snakes,shapes,decorations.fractals,decorations.pathmorphing,patterns,decorations.markings}

\title{
Mixed boundary conditions in
AdS$_2$/CFT$_1$ \\
from the coupling with a Kalb-Ramond field 
}

\author{Diego H. Correa, Maximiliano G. Ferro and Victor I. Giraldo-Rivera}

\affil{\it {Instituto de F\'isica La Plata - CONICET} \&\\
{\it Departamento de F\'\i sica,  Universidad Nacional de La Plata}\\ {\it C.C. 67, 1900,  La Plata, Argentina}}

\begin{document}

\maketitle

%=========================================

\vspace{0.4cm}
\begin{center}{\bf Abstract}
\end{center}

\begin{minipage}{14cm}
The open string dual to a 1/6 BPS Wilson line in the ${\cal N} = 6$ super Chern-Simons-matter theory is coupled to a flat Kalb-Ramond field. We show that the resulting boundary term  imposes mixed boundary conditions on the fields that describe the fluctuations on the world-sheet. These boundary conditions fix a combination of the derivatives of the fluctuations, parallel and transverse to the boundary. We holographically compute the correlation functions of insertions on the Wilson line in terms of world-sheet Witten diagrams. We observe that their functional dependence is consistent with the conformal symmetry on the line.
\end{minipage}

\section{Introduction}
The dynamics of fields propagating within Anti-de Sitter (AdS) space is known to encode the set of correlation functions of certain conformal field theories (CFTs) \cite{malda1998,gkp1998,witten1998}. 
The presence of a boundary in AdS implies that different boundary conditions for fields give rise to diverse behaviors and properties in the corresponding dual CFT. For massive scalar fields, there exists a range of masses for which  two different scaling dimensions are admissible for the dual primary operators, associated with either imposing the so-called {\it regular} or {\it alternate} boundary conditions \cite{Klebanov:1999tb}. Additionally, within the same mass range, it is also possible to impose certain mixed boundary conditions that are related to field theory operators interpolating between primaries of different scaling dimensions. These mixed boundary conditions are interpreted in terms of flows of the renormalization group \cite{Witten:2001ua,GM,GK,Hartman:2006dy}.

For massless scalar fields in AdS$_2$, a different type of boundary conditions that mix longitudinal and transverse derivatives was considered in \cite{CGRS}. Since the mixing parameter is dimensionless in that case, one might expect these boundary conditions to be associated with a marginal deformation rather than with a flow of the renormalization group. However, determining whether these boundary conditions correspond to a conformal theory on the line or not, is not straightforward, as this will depend on the actual details of the AdS$_2$/CFT$_1$ realization. 

Massless scalar fields in AdS$_2$ typically arise when studying the fluctuations on an open string world-sheet dual to line operators in various $d$-dimensional CFTs. An interesting case is that of the Wilson loops in the ABJ(M) model, a prototypical example of the AdS/CFT correspondence where the ${\cal N} = 6$ super Chern-Simons theory  with gauge group $U(N)_k\times U(N)_{-k}$ is conjectured to be equivalent to type IIA string theory in AdS$_4 \times \mathbb{CP}^3$ \cite{Aharony:2008ug}. 

This model admits a simple and interesting generalization \cite{Aharony:2008gk}, in which the gauge group in the Chern-Simons theory is taken to be $U(N+\ell)_k\times U(N)_{-k}$. 
In this case, the dual string theory description includes an additional  flat Kalb-Ramond field, having a non-trivial holonomy on  the non-contractible $\mathbb{CP}^1\subset\mathbb{CP}^3$. Being the Kalb-Ramond field flat, its coupling to the open string only leads to a boundary term. As we will see in this article, the boundary term from the coupling with the Kalb-Ramond field is responsible for the materialization of the aforementioned kind of mixed boundary conditions when we study the open string dual to the 1/6 BPS {\it bosonic} Wilson line \cite{Drukker:2008zx,Chen:2008bp,Rey:2008bh}.

The analysis of Witten diagrams for the fluctuations on an AdS$_2$ world-sheet enables the holographic computation of the correlation functions of some  operators ${\cal O}(t)$ on a line. These line correlators are related to the expectation values of local operator insertions along a Wilson line \cite{Drukker:2006xg,Giombi:2017cqn},
\begin{align}
    \langle\!\langle
{\cal O}(t_1)\cdots{\cal O}(t_n)
\rangle\!\rangle =
\langle {\rm tr}{\,\cal P}\left[{\cal O}(t_1)\cdots{\cal O}(t_n) \, W_{1/6}\right] 
\rangle\,.
\end{align}
This effective theory on a line can preserve conformal symmetry at the quantum level or not, which should be reflected in the functional dependence of the correlation functions of excitations. 

The main objective of our work is to test perturbatively, in inverse powers of the 't Hooft coupling, the conformal covariance of the theory on the line dual to the open string with mixed boundary conditions. In order to do that, we will analyze the functional dependence of holographic 4-point functions. For example, conformal symmetry on the line would require that the 4-point correlator of a primary field of weight $\Delta$ has to be of the form
\begin{align}
    \langle\!\langle
{\cal O}(t_1){\cal O}(t_2){\cal O}(t_3){\cal O}(t_4)
\rangle\!\rangle = 
\frac{G(u)}{(t_2-t_1)^{2\Delta}(t_4-t_3)^{2\Delta}}\,,
\qquad \quad {\rm for}\quad u=\frac{(t_1-t_2)(t_3-t_4)}{(t_1-t_3)(t_2-t_4)}\,.
\end{align}
We will consider this for the concrete setup that corresponds to insertions in the 1/6 BPS bosonic Wilson line in the ABJ(M) model.

The paper is organized as follows. In section \ref{openstring} we introduce the open string setup dual to Wilson loops in the ABJ(M) model and the possible boundary conditions on the world-sheet excitations. We show that the coupling with a flat Kalb-Ramond field leads to boundary conditions combining longitudinal and transverse derivatives and determine the propagators corresponding to them. Section \ref{correlators} is dedicated to the holographic computation of 2-point and 4-point correlation functions using Witten diagrams and the propagators associated with mixed boundary conditions. Finally, in section \ref{discu} we conclude with a discussion of our results. In the appendix \ref{oneloop} we provide details about the self-energy diagrams contributing to the 1-loop correction of the propagators.

\section{Open strings in AdS$_4 \times \mathbb{CP}^3$}
\label{openstring}
We shall consider open strings embedded  in AdS$_4 \times \mathbb{CP}^3$, whose metric can be written as
\begin{equation}
    ds^2 = L^2\left(ds^2_{\rm AdS_4} + 4 ds^2_{\mathbb{CP}^3}   \right)\,,
\end{equation}
where $L$ is the AdS$_4$ radius. The type IIA fields supporting this geometry are
\begin{equation}
  e^{\phi} = \frac{2L}{k}  \,,
  \quad
  F^{(4)} = \frac{3}{2}k L^2 \text{vol}({\rm AdS}_4)\,,
  \quad
  F^{(2)} = \frac{k}{4} dA\,,
\end{equation}
where $k$ is an integer identified with the Chern-Simons level in the dual field theory, which has gauge group
$U(N)_k\times U(N)_{-k}$ \cite{Aharony:2008ug}.
Defining the 't Hooft coupling as $\lambda = 2\pi^2 N/k$, its relation to the AdS$_4$ radius is
$L^2 = \sqrt{\lambda}\alpha'$ \footnote{This is not the most usual convention in ABJM, but it leads to an effective string tension $T=\frac{L^2}{2\pi\alpha'}=\frac{\sqrt\lambda}{2\pi}$, which will facilitate the comparison with the ${\cal N}=4$ SYM results of \cite{Beccaria:2019dws} in the limit of Neumann boundary conditions.}.
The 2-form $dA$ is proportional to the K\"ahler form of the $\mathbb{CP}^3$. In angular coordinates one has
\begin{eqnarray}
    ds_{\mathbb{CP}^3}^2 &=& \frac{1}{4}
\left[ d\chi^2 + \cos^2\frac{\chi}{2}\left(d \theta_1^2+\sin^2 \theta_1 d \varphi_1^2\right)
+\sin^2\frac{\chi}{2}\left(d \theta_2^2 
+\sin ^2 \theta_2 d \varphi_2^2\right)
\right. \nonumber
\\
&& + \left .\sin ^2 \frac{\chi}{2} \cos ^2 \frac{\chi}{2}\left(d \xi+\cos \theta_1 d \varphi_1-\cos \theta_2 d \varphi_2\right)^2\right] \,,
\label{cp3metric}
\end{eqnarray} 
with ranges $0 \leq \chi, \theta_1, \theta_2 \leq \pi$, $0 \leq \varphi_1, \varphi_2 \leq 2\pi$  and $0 \leq \xi \leq 4\pi$ and
\begin{equation}
    A = \cos\chi d\xi + 2 \cos^2\frac{\chi}{2}\cos\theta_1 d\varphi_1
    + 2 \sin^2\frac{\chi}{2}\cos\theta_2 d\varphi_2\,.
\end{equation}

In this background we will consider an open string ending along a straight line at the boundary of AdS$_4$, whose classical world-sheet is AdS$_2$. This gives the holographic dual description of a supersymmetric straight Wilson line in the ${\cal N}=6$ super Chern-Simons theory. It is convenient to write the (Euclidean) AdS$_4$ metric with an explicit (Euclidean) AdS$_2$ foliation
\cite{Giombi:2017cqn}
\begin{equation}
ds^2_{\rm AdS_4} = \frac{\left(1+\frac14x^2\right)^2}{\left(1-\frac14x^2\right)^2}\left(\frac{dt^2+dz^2}{z^2}\right)+ \frac{dx^i dx^i}{\left(1-\frac14x^2\right)^2}\,.
\end{equation}
The metric of the $\mathbb{CP}^3$ can be  parameterized alternatively in terms of 3 complex coordinates
\begin{equation}
ds^2_{\mathbb{CP}^3} = \frac{d\bar w_a dw^a}{1+|w|^2} - \frac{d\bar w_a w^a dw^b \bar w_b }{\left(1+|w|^2\right)^2}\,.
\end{equation}

In the static gauge $t=\tau$ and $z=\sigma$, the Nambu-Goto action becomes
\begin{eqnarray}
    S_{\rm NG} \! = \! 
    \frac{\sqrt\lambda}{2\pi}
    \int                        \!d^2\sigma 
    %&& \!\!\!\!\!\!\!\!\!\!\!
    %\left\
    \sqrt{
    {\rm det}\!\left[\frac{\left(1+\frac14x^2\right)^2}{\left(1-\frac14x^2\right)^2}g_{\mu\nu}
    +  \frac{\partial_\mu x^i \partial_\nu x^i}{\left(1-\frac14x^2\right)^2}    +
    \frac{4 \partial_\mu \bar w_a \partial_\nu w^a}{1+|w|^2} - \frac{4 \partial_\mu \bar w_a w^a \partial_\nu w^b \bar w_b }{\left(1+|w|^2\right)^2}
\right]}
\label{NGaction}
\end{eqnarray}
where $g_{\mu\nu}$ is the metric of (Euclidean) AdS$_2$ 
\begin{equation}
ds^2_{\rm AdS_2} = \frac{dt^2+dz^2}{z^2}\,.
\end{equation}

Expanding \eqref{NGaction}  in powers of $x^i$ and $w^a$, one gets the action for 2 real scalars of mass $m^2 = 2/L^2$ and 3 complex massless scalars respectively \cite{Bianchi:2020hsz}.
The dual gauge theory admits a variety of supersymmetric Wilson loops. There is a 1/6 BPS Wilson loop that involves only bosonic fields \cite{Drukker:2008zx,Chen:2008bp,Rey:2008bh}, a 1/2 BPS one \cite{Drukker:2009hy} and even a family that interpolates between them \cite{Ouyang:2015iza,Ouyang:2015bmy,Castiglioni:2022yes,Castiglioni:2023uus}
(see \cite{Drukker:2019bev} for a review). These Wilson loops differ not only in the amount of supersymmetry preserved but also in their bosonic R-symmetries. For example, the symmetries of the 1/2 BPS and the 1/6 BPS bosonic Wilson loops are $SU(3)$ and $SU(2)\times SU(2)$ respectively. In the open string theory description, this translates into different boundary conditions on the $\mathbb{CP}^3$ coordinates. To describe the 1/2 BPS Wilson line, Dirichlet boundary conditions must be imposed on all $\mathbb{CP}^3$ coordinates. However, to describe the 1/6 BPS bosonic Wilson line, the string has to be uniformly smeared over a $\mathbb{CP}^1\subset\mathbb{CP}^3$ \cite{Drukker:2008zx}, which can be achieved by imposing Neumann boundary conditions on 2 of the $\mathbb{CP}^3$ coordinates and Dirichlet on the rest. With this in mind, we will focus on the coordinates along a particular $\mathbb{CP}^1\subset\mathbb{CP}^3$, which, for the sake of definiteness, is taken to be the one given by $\chi = 0$, {\it i.e.} the one parametrized by $\theta_1$ and $\varphi_1$. We can also use embedding $\{Y^A\}$, with ${A=1,2,3}$, to describe the  $\mathbb{CP}^1$ as a surface in $\mathbb{R}^3$ imposing $\delta^{AB} Y^A\, Y^B =1$. 
Ignoring all the other transverse fluctuations, we obtain the following action for the $\mathbb{CP}^1$ embedding coordinates
\begin{eqnarray}
\label{NGCP1}
    S_{\mathbb{CP}^1}  =  \frac{\sqrt\lambda}{2\pi}\int d^2\sigma \sqrt{
    {\rm det} 
    \left( g_{\mu\nu} + \partial_\mu Y^A \partial_\nu Y^A \right)}\,.
\end{eqnarray}

The $\mathbb{CP}^1$ fluctuations can be taken around a fix position $n^A$. In terms of the orthogonal part to it,  one has
\begin{equation}
\label{eq:coordexpand}
    Y^A = n^A \sqrt{1-\zeta^2} + \zeta^A\,,
\end{equation}
with $n^A\cdot \zeta^A=0$ \cite{Beccaria:2019dws}. Thus, for small $\zeta^A = \tfrac{\sqrt{2\pi}}{\lambda^{1/4}} y^A$
\begin{eqnarray}
     Y^A= n^A + \frac{\sqrt{2\pi}}{\lambda^{1/4}} y^A -\frac{\pi}{\sqrt{\lambda}} y^2 n^A + {\cal O}(\tfrac{1}{\lambda})\,,
     \label{Yexpa}
\end{eqnarray}
and replacing in \eqref{NGCP1}, we obtain
\begin{eqnarray}
\label{NGCP1expa}
    S_{\mathbb{CP}^1} \!\!& = &\!\! \ \int d^2\sigma   {\sqrt{g}}
    \left[\tfrac{1}{2}\partial_\mu y^A
    \partial^\mu y^A +
    \frac{\pi}{\sqrt \lambda} y^A y^B \partial_\mu y^A
    \partial^\mu y^B
    +
    \frac{\pi}{4\sqrt \lambda}(\partial_\mu y^A
    \partial^\mu y^A)^2 
    \right.
    \nonumber
    \\
    && \left. \qquad \qquad -
     \frac{\pi}{2\sqrt \lambda}\partial_\mu y^A
    \partial^\mu y^B \partial_\nu y^A
    \partial^\nu y^B+ {\cal O}(\tfrac{1}{\lambda})
    \right]\,.
\end{eqnarray}

This quartic action gives the vertices that 
play an important role in the holographic computation of 4-point correlation functions of excitations along the dual line.

\subsection{Dirichlet, Neumann and mixed boundary conditions}
Before to proceed, let us make a few comments about the boundary conditions used to specify the AdS$_2$ propagators and their relation to the variational problem. The on-shell variation of the action \eqref{NGCP1expa} is, up to quadratic order, a boundary term
\begin{equation}
\label{variation S0}
	\delta S_{\mathbb{CP}^1}^{(2)} = -
 \int_{-\infty}^{\infty}dt \ \partial_z {y^A}\delta y^A \big|_{z=0}\,,
\end{equation}
whose vanishing would require to set Dirichlet boundary conditions, {\it i.e.} $\delta y^A \big|_{z=0} = 0$. The   imposition of Neumann boundary conditions 
would be required when another boundary term is added to the quadratic action
\begin{equation}
    \tilde{S}_{\mathbb{CP}^1}^{(2)} = S_{\mathbb{CP}^1}^{(2)} + 
    \int_{-\infty}^{\infty}dt \ \partial_z {y^A} y^A \big|_{z=0}\,,
\end{equation}
so that its on-shell variation becomes
\begin{equation}
\label{variation tildeS0}
	\delta \tilde{S}_{\mathbb{CP}^1}^{(2)} = 
 \int_{-\infty}^{\infty}dt \ \delta(\partial_z {y^A})y^A \big|_{z=0}\,,
\end{equation}
whose vanishing would be achieved by demanding $\delta(\partial_z {y^A}) \big|_{z=0} = 0$.

In the ABJ generalization, the gauge group is
$U(N+\ell)_k\times U(N)_{-k}$. This difference in the ranks of the gauge group factors is accounted, in the dual string theory, by incorporating a flat Kalb-Ramond field with a non-trivial holonomy on a $\mathbb{CP}^1$ \cite{Aharony:2008gk}
\footnote{This $\mathbb{CP}^1$ is not the one specified before by $\chi=0$ but the non-contractible 2-cycle of the $\mathbb{CP}^3$.}.
\begin{equation}
    \frac{1}{2\pi}\int_{\mathbb{CP}^1} B^{(2)} = \frac{\ell}{k}\,,
    \qquad
   B^{(2)} = \frac{\cal B}{2} dA\,.
\end{equation}
Unitarity requires that $|\ell|\leq k$ \cite{Aharony:2008gk}.

Since the Kalb-Ramond field is flat, its coupling to the open string adds just a boundary term. While the equations of motion remain unaffected, the boundary conditions on the $\mathbb{CP}^3$ coordinates can change. To analyze this effect let us consider the expansion of the Kalb-Ramond term
\begin{eqnarray}
 \!\!\!\! \frac{1}{2}\frac{\sqrt\lambda}{2\pi}  B_{\mu\nu}\epsilon^{\alpha\beta}
\partial_\alpha X^\mu \partial_\beta X^\nu
\!\!\!\! %&=& \!\!\!\!- \frac{\sqrt\lambda}{2\pi}{\cal B}
%\sin \theta_1 \epsilon^{\alpha\beta} \partial_\alpha \theta_1 \partial_\beta \varphi_1\nonumber\\
%\!\!\!\! 
&=& \!\!\!\! - \frac{\sqrt\lambda}{2\pi}{\cal B}\epsilon^{ABC}Y^A\partial_\tau Y^B \partial_\sigma Y^C
\nonumber\\
\label{KRterm}
\!\!\!\! &=& \!\!\!\! -{\cal B} \epsilon^{ABC}
\left(n^A \partial_z y^B \partial_t y^C 
-  \frac{\pi}{\sqrt\lambda} n^A y^2 \partial_z y^B \partial_t y^C  \right.
\\
&& \left.\qquad - \frac{2\pi}{\sqrt\lambda} n^C y^D \partial_t y^D y^A \partial_z y^B - \frac{2\pi}{\sqrt\lambda} n^B y^A y^D \partial_z y^D \partial_t y^C +{\cal O}(\tfrac{1}{\lambda})\right)\,.\nonumber 
\end{eqnarray}
Thus, up to quadratic order
\begin{eqnarray}
\label{NGCP1expaplusB}
    S_{\mathbb{CP}^1}^{(2)} \!\!& = &\!\! \int d^2\sigma  
    \left(\sqrt{g} \tfrac{1}{2}  \partial_\mu y^A
 \partial^\mu y^A 
 -{\cal B}\epsilon^{ABC}
n^A \partial_z y^B \partial_t y^C 
 \right)\,,
 \end{eqnarray}
whose on-shell variation becomes
\begin{equation}
\label{variation S0plusB}
	\delta S_{\mathbb{CP}^1}^{(2)} = -\int_{-\infty}^{\infty} dt \left( \partial_z {y^A}
 +{\cal B}
 \epsilon^{ABC} n^B
\partial_t y^C
 \right)\delta y^A \big|_{z=0}\,.
\end{equation}

Thus, the boundary term from the coupling to the Kalb-Ramond field does not change the boundary conditions in this case, as still $\delta y^A \big|_{z=0} = 0$ is required. However, when the same term is added to the quadratic action which originally had  Neumann boundary conditions, we get
\begin{equation}
\label{variation tilde S0plusB}
	\delta \tilde{S}_{\mathbb{CP}^1}^{(2)} = \int_{-\infty}^{\infty}dt \ \delta
 \left(\partial_z {y^A}
 -{\cal B}
\epsilon^{ABC}n^B\partial_t y^C  \right) y^A \big|_{z=0}\,.
\end{equation}
Therefore, in this other case the boundary conditions do get modified to
\begin{equation}
     \delta
 \left(\partial_z {y^A}
 -{\cal B}
\epsilon^{ABC}n^B\partial_t y^C  \right)\big|_{z=0} = 0\,.
\label{eq:BC}
\end{equation}

As seen in \eqref{KRterm}, the coupling with the Kalb-Ramond field also gives rise to terms of quartic order in the orthogonal fluctuations, which will be central in the holographic computation of the 4-point functions.
Using the orthogonality condition $n^A y^A =0$, it is not difficult to show that these quartic terms are
\begin{eqnarray}
    {\sf KR}^{(4)}&=& \mathcal{B} \epsilon^{ABC}\left( \tfrac{1}{2} n^A y^2 \partial_z y^B \partial_t y^C  + n^C y^D \partial_t y^D y^A \partial_z y^B + n^B y^A y^D \partial_z y^D\partial_t y^C \right)\nonumber\\
    &=&- \mathcal{B} \epsilon^{ABC} \tfrac{1}{2} n^A y^2 \partial_z y^B \partial_t ^C \,.
\end{eqnarray}
Finally, and after a few successive integrations by parts, we obtain
\begin{eqnarray}
{\sf KR}^{(4)}
 &=& -\frac{\mathcal{B}}{8}\Big(\partial_z\left(\epsilon^{ABC}n^A y^2 y^B \partial_t y^C \right)-\partial_t\left(\epsilon^{ABC}n^A y^2 y^B \partial_z y^C \right)\Big)\,.
\end{eqnarray}

\subsection{AdS$_2$ propagators for mixed boundary conditions}

We would like to address the problem of quadratic fluctuations with specific mixed boundary
conditions \eqref{eq:BC}, 
\begin{eqnarray}
\label{eq:quadraticproblem}
   \Box  y^A = 0\,,\qquad
      \left.\left( \partial_z y^A-{\cal B}\, \epsilon^{ABC}n^B\partial_t y^C \right)\right|_{z = 0}=J^A(t)\,,
\end{eqnarray}
by using a suitable propagator. The free propagator for massless scalar fields in AdS$_2$ is given in terms of a Green's function
\begin{equation}
\langle y^A(\sigma) y^B(\sigma') \rangle_{{0},n} =  -{\rm G}^{AB}(\sigma,\sigma')\,,
\label{propadef}
\end{equation}
which is a solution of 
\begin{equation}
\label{boxG}
\Box  {\rm G}^{AB} =  (\delta^{AB}-n^A n^B)\delta(t-t')\delta(z-z')\,. 
\end{equation}
The additional sub-index $n$ in \eqref{propadef} serves to emphasize that the propagator depends on the values $n^A$ chosen to compute the fluctuations. These boundary conditions are proposed to describe the string dual to the 1/6 BPS bosonic Wilson line, which is uniformly smeared over a $\mathbb{CP}^1$. Thus, we would have to average over the possible values of $n^A$ eventually.

To determine the Green's function suitable for the boundary conditions \eqref{eq:BC}, we can use the Green's third identity
\begin{equation}
    y^A(z',t') = \int_{-\infty}^{\infty}\! dt \left.\left( {\rm G}^{AB}(z,t,z',t')\partial_z y^B(z,t)-
    \partial_z {\rm G}^{AB}(z,t,z',t')y^B(z,t) 
    \right)\right|_{z=0}
\end{equation}
and rewrite it by adding the integral of a $t$-derivative. We get
\begin{align}
y^A(z',t') = &\int_{-\infty}^{\infty}\! dt \left\{ {\rm G}^{AB}(z,t,z',t') \left( \partial_z y^B(z,t) -{\cal B} \,\epsilon^{BCD}n^C\partial_t y^D \right)\right.
\nonumber\\
 &\qquad\qquad -\left.\left.\left(  \partial_z {\rm G}^{AD}(z,t,z',t') 
  + {\cal B}\, \partial_t {\rm G}^{AB}(z,t,z',t') \epsilon^{BCD}n^C\right) y^D
    \right\}\right|_{z=0}
\end{align} 
Thus, imposing the following boundary condition for the Green's function
\begin{equation}
\label{bcG}
    \left. \left(\partial_z {\rm G}^{AD}(z,t,z',t') 
  + {\cal B}\, \partial_t {\rm G}^{AB}(z,t,z',t') \epsilon^{BCD}n^C \right)\right|_{z=0} = 0\,,
\end{equation}
we can easily write a solution to the problem \eqref{eq:quadraticproblem} as
\begin{align}
y^A(z',t') 
& = \int_{-\infty}^{\infty}\! dt\, \left.\left[ {\rm K}^{AB}(t,z',t') \left( \partial_z y^B(z,t) - {\cal B} \,\epsilon^{BCD}n^C\partial_t y^D \right)\right]\right|_{z=0}
\end{align} 
where we have introduced the bulk-to-boundary propagator, obtained in this case as
\begin{equation}
  {\rm K}^{AB}(t,z',t') = \lim_{z\to0}  {\rm G}^{AB}(z,t,z',t')\,.
\end{equation}
At certain point, we will also consider
the boundary-to-boundary propagator, 
\begin{equation}
  {\rm D}^{AB}(t,t') = \lim_{z\to0}
  \lim_{z'\to0}{\rm G}^{AB}(z,t,z',t')\,.
\end{equation}

To proceed, we need a Green's function that satisfies \eqref{boxG} and \eqref{bcG}. It is convenient to split ${\rm G}^{AB}$ into symmetric and antisymmetric parts as
\begin{equation}
    {\rm G}^{AB}(\sigma,\sigma')
    =
    (\delta^{AB}-n^A n^B) {\rm G}_{\sf s}(\sigma,\sigma')
    +\epsilon^{ABC}n^C {\rm G}_{\sf a}(\sigma,\sigma')
    \,.
\end{equation}
The antisymmetric term in the propagator of the string coordinates arises from coupling the open string with the Kalb-Ramond field \cite{Chu:1998qz,Schomerus:1999ug,Seiberg:1999vs}. 

It is not difficult to see that the Green's function suitable for our problem is given by
\begin{align}
{\rm G}_{\sf s}(\sigma,\sigma')
    &= 
    \frac{1}{1-{\cal B}^2}{\rm G}_N(\sigma,\sigma')-\frac{{\cal B}^2}{1-
    {\cal B}^2}{\rm G}_D(\sigma,\sigma')\,, \\
{\rm G}_{\sf a}(\sigma,\sigma')
    &=  \frac{{\cal B}}{1-{\cal B}^2} {\rm G}_{0}(\sigma,\sigma')
    \,,
\end{align}
where 
\begin{align}\label{greenfunctions}
    {\rm G}_N(\sigma,\sigma') &=  \frac{1}{4\pi}\left(
    \log\left((t-t')^2+ (z-z')^2 \right)
    +\log\left((t-t')^2+ (z+z')^2 \right)
    \right)\,,
    \\
    {\rm G}_D(\sigma,\sigma') &=  \frac{1}{4\pi}\left(
    \log\left((t-t')^2+ (z-z')^2 \right)
    -  \log\left((t-t')^2+ (z+z')^2 \right)
    \right)\,,
    \\
    {\rm G}_0(\sigma,\sigma') &= -\frac{1}{\pi}\tan^{-1}
    \left(\frac{t-t'}{z+z'}\right)\,.
\end{align}

As expected, in the limit ${\cal B}\to 0$, we recover the  Green's function corresponding to Neumann boundary conditions. For  the bulk-to-boundary propagator we have
\begin{equation}
    {\rm K}^{AB}(z,t;t')= \frac{1}{1-{\cal B}^2} (\delta^{AB}-n^A n^B) {\rm K}_N(z,t;t') +  \frac{\cal B}{1-{\cal B}^2} \epsilon^{ABC} n^C {\rm K}_0(z,t;t')\,,
\end{equation}
with
\begin{align}
{\rm K}_N(z,t;t') &=\frac{1}{2\pi}\log\left((t-t')^2+z^2\right)\,,
\\
{\rm K}_0(z,t;t') &=
 -\frac{1}{\pi}\tan^{-1}
    \left(\frac{t-t'}{z}\right)\,.
\end{align}
Notice that
\begin{equation}
 \partial_t {\rm K}_N=-\partial_z {\rm K}_0\,,
\end{equation}
and the following boundary conditions are satisfied
\begin{eqnarray}  
\lim_{z\rightarrow 0}\partial_z {\rm K}_N(z,t;t')=\delta(t-t')\,,\qquad  \lim_{z\rightarrow 0}\partial_t {\rm K}_0(z,t;t')=\delta(t-t')\,.
\end{eqnarray}

\section{Correlators on the line from Witten diagrams on $AdS_2$}
\label{correlators}

The embedding coordinates $Y^A$ can be put in correspondence with scalar operators inserted along the Wilson line. These scalar operators should be arranged in a triplet of the $SU(2)\subset SU(4)$. Such triplet can be constructed using bilinears of scalars fields of the ABJ(M) theory and they are non-protected: while their scaling dimension is 1 at zero coupling, it diminishes to zero in the leading strong coupling limit. In the following, we will study correlations functions of the operators dual to embedding coordinates $Y^A$. We will begin with 2-point functions to read from them  $1/\sqrt{\lambda}$ corrections to the vanishing scaling dimension. Then, we will study 4-point functions to evaluate the conformal symmetry of the theory defined by this line defect.

\subsection{2-point functions}
\label{2point}
The 1-dimensional conformal symmetry should fix the form of the 2-point correlator of embedding coordinates to be
\begin{align}
 \langle Y^A\left(t_1\right) Y^B\left(t_2\right)\rangle &=
  \frac{C_Y\delta^{AB}}{(t_2-t_1)^{2\Delta}}
  = C_Y\delta^{AB}
  \left[1-\Delta\ {\rm L}_{12} +\tfrac{1}{2}\Delta^2 {\rm L}_{12}^2 +\cdots\right]\,,
 \end{align}
where 
\begin{equation}
{\rm L}_{ij} :=  \log\left((t_i-t_j)^2\right)\,.
\end{equation}
The scaling dimension $\Delta$ and normalization $C_Y$ can be expanded at strong coupling as
\begin{equation}
    \Delta =  \tfrac{d_1}{\sqrt\lambda}+\tfrac{d_2}{\lambda}+\cdots\,,\qquad
    C_Y = c_0+\tfrac{c_1}{\sqrt\lambda}+\tfrac{c_2}{\lambda}+\cdots\,.
\label{DeltaCexpansions}
\end{equation}
Thus,
\begin{align}
 \langle Y^A\left(t_1\right) Y^B\left(t_2\right)\rangle = &
  \delta^{AB} c_0 + \frac{\delta^{AB}}{\sqrt{\lambda}}
  \left(c_1- d_1 {\rm L}_{12}\right)
   \nonumber\\
   & + \frac{\delta^{AB}}{{\lambda}}
  \left(c_2 - (c_0 d_2+c_1 d_1) {\rm L}_{12} +\tfrac{1}{2} c_0 d_1^2 {\rm L}_{12}^2 \right)
  + \mathcal{O}(\tfrac{1}{\lambda^{3/2}})
  \,.
  \label{2ptexpa}
   \end{align}
The successive orders $c_i$ and $d_i$ can be derived from the strong coupling expansion of this 2-point function. Using the expansion of the embedding coordinates into fluctuations around a fixed $n^A$ \eqref{Yexpa}, the leading order of the correlation function between two embedding coordinates becomes
\begin{align}
 &\langle Y^A\left(t_1\right) Y^B\left(t_2\right)\rangle= 
  \left\langle n^A n^B \right\rangle + 
  \frac{2\pi}{\sqrt\lambda}\left
  \langle y^A(t_1) y^B(t_2) \right\rangle_{0} + \mathcal{O}(\tfrac{1}{\lambda}) \,.
 \end{align}
Omitting the sub-index $n$ in the brackets means that we are also averaging over $n^A$. Some useful averages are the following,
\begin{align}
    \langle n^A \rangle &= 0\,,
    \hspace{1.27cm}
    \langle n^A n^B \rangle = \frac{1}{3}\delta^{AB}\,,
    \\
    \langle n^A n^B n^C \rangle &= 0\,,
        \quad
    \langle n^A n^B n^C n^D \rangle = \frac{1}{15}\left(\delta^{AB}\delta^{CD}+\delta^{AC}\delta^{BD}+\delta^{AD}\delta^{CB}\right)\,.
\end{align}
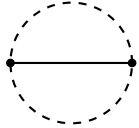
\begin{figure}[h!]
    \centering
    \begin{tikzpicture}
    \draw[dashed,thick] (0,0) circle [radius=0.8cm];
    \draw[thick,cap=round]  (-0.8,0) -- (0.8,0);
    \draw[fill=black] (0.8,0) circle[radius=0.05cm];
    \draw[fill=black] (-0.8,0) circle[radius=0.05cm];
    \end{tikzpicture}
    \caption{Contribution order $\tfrac{1}{\sqrt\lambda}$ to $\langle Y^A(t_1)Y^B(t_2)\rangle$}
    \label{fig:2pt2}
\end{figure}

Using that the propagator \eqref{propadef} in the boundary-to-boundary limit becomes
\begin{equation}
{\rm D}^{AB}(t_1,t_2)= \frac{1}{2\pi}\frac{1}{1-{\cal B}^2} (\delta^{AB}-n^A n^B) \log\left((t_1-t_2)^2\right) +  \frac{1}{2}\frac{\cal B}{1-{\cal B}^2} \epsilon^{ABC} n^C \text{sign}(t_2-t_1)\,,
\end{equation}
we get 
 \begin{align}
   \langle Y^A\left(t_1\right) Y^B\left(t_2\right)\rangle& = \frac{1}{3}\delta^{AB}\left[1 -  \frac{2}{\sqrt\lambda}  \frac{1}{1-{\cal B}^2}\log\left((t_1-t_2)^2\right) + {\cal O}(\tfrac{1}{\lambda})\right]\,.
\end{align}
From this we can read that $d_1 =  \frac{2}{1-{\cal B}^2}$, $c_0=\frac{1}{3}$ and $c_1=0$. In other words,
\begin{align}
    \Delta =  \frac{2}{\sqrt\lambda(1-{\cal B}^2)} +{\cal O}(\tfrac{1}{\lambda})    \,,\qquad
    C_Y = \frac{1}{3} + {\cal O}(\tfrac{1}{\lambda})\,. 
\end{align}

To proceed beyond, we need to evaluate the 2-point function of embedding coordinates keeping track of terms order $\frac{1}{\lambda}$. For this we have to go further in the expansion into fluctuations around $n^A$ \eqref{Yexpa} and also to include 1-loop corrections to the propagators,
\begin{align}
\label{YYnext}
\langle Y^A\left(t_1\right) Y^B\left(t_2\right)\rangle = & \frac{1}{3}\delta^{AB}\left[1 -  \frac{2}{\sqrt\lambda}  \frac{1}{1-{\cal B}^2}\log\left((t_1-t_2)^2\right) \right] 
\\
 & +\frac{\pi^2}{\lambda} \langle
 n^A n^B y^C(t_1)y^C(t_1) y^D(t_2)y^D(t_2) 
 \rangle_0
 +
 \frac{2\pi}{\lambda}\left
  \langle y^A(t_1) y^B(t_2) \right\rangle_{1}
  + {\cal O}(\tfrac{1}{\lambda^{3/2}})\,.
  \nonumber
\end{align}
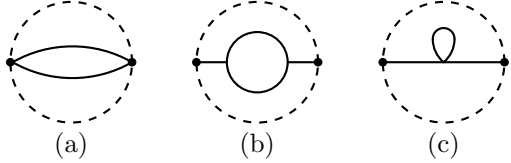
\begin{figure}[h!]
    \centering
    \begin{tikzpicture}
    \draw[dashed,thick] (0,0) circle [radius=0.8cm];
    \draw[thick,cap=round] (0.8,0) arc (60:120:1.57cm);
    \draw[thick,cap=round] (0.8,0) arc (-60:-120:1.57cm);
    \draw[fill=black] (0.8,0) circle[radius=0.05cm];
    \draw[fill=black] (-0.8,0) circle[radius=0.05cm];
    \node at (0,-1.1) {(a)};
    \end{tikzpicture}
    \hspace{0.5cm}
    \begin{tikzpicture}
    \draw[dashed,thick] (0,0) circle [radius=0.8cm];
    \draw[thick,cap=round]  (-0.8,0) -- (0.8,0);
    \draw[fill=white,thick] (0,0) circle [radius=0.4cm];
    \draw[fill=black] (0.8,0) circle[radius=0.05cm];
    \draw[fill=black] (-0.8,0) circle[radius=0.05cm];
    \node at (0,-1.1) {(b)};
    \end{tikzpicture}
    \hspace{0.5cm}
    \begin{tikzpicture}
    \draw[dashed,thick] (0,0) circle [radius=0.8cm];
    \draw[thick,cap=round]  (-0.8,0) -- (0.8,0);
    \draw[fill=black] (0.8,0) circle[radius=0.05cm];
    \draw[fill=black] (-0.8,0) circle[radius=0.05cm];
    \draw[thick,cap=round] (0,0) arc (-50:-0:0.4cm);
    \draw[thick,cap=round] (0,0) arc (230:180:0.4cm);
    \draw[thick,cap=round] (0.143,0.31) arc (0:180:0.143cm);
    \node at (0,-1.1) {(c)};
    \end{tikzpicture}    
    \caption{Contributions order $\tfrac{1}{\lambda}$ to $\langle Y^A(t_1)Y^B(t_2)\rangle$.}
    \label{fig:2pt2}
\end{figure}
The Witten diagrams representing the terms of the second line of \eqref{YYnext} are depicted in Fig. \ref{fig:2pt2}.
The first term is represented by the diagram 2(a) and its computation is rather straightforward. Before averaging over $n^A$
\begin{align}
2{\rm (a)}_n &= 2 \times \frac{\pi^2}{\lambda} 
 n^A n^B 
\langle y^C(t_1) y^D(t_2)\rangle_{0,n} \langle y^C(t_1) y^D(t_2)\rangle_{0,n} \nonumber
 \\
 & = \frac{1}{\lambda}  n^A n^B\frac{1}{(1-{\cal B}^2)^2}
 \left({\rm L}_{12}^2+\pi^2{\cal B}^2\right)\,.
\end{align}

Diagrams 2(b) and 2(c) come from the 1-loop corrections to the propagator 
\begin{align}
2{\rm (b)}_n+2{\rm (c)}_n 
& = 
\frac{2\pi}{\lambda} \langle y^A(t_1) y^B(t_2)\rangle_{1,n}\,.
\end{align}

It is possible to argue (see appendix \ref{oneloop}) that these 1-loop corrections will be of the form 
\begin{align}
\langle y^A(t_1) y^B(t_2)\rangle_{1,n}
  = &
 {\left(\delta^{AB}- n^A n^B \right)}
 \left(
 f_0+ f_1 {\rm L}_{12} + 
      f_2 {\rm L}_{12}^2\right)
      \nonumber\\& 
      +
 {\epsilon^{ABC}n^C}
 {\rm sign}(t_1-t_2)\left(
 g_0  + g_1 {\rm L}_{12}\right)\,,
 \label{2pt1loop}
\end{align}
for some constant coefficients $f_0, f_1, f_2, g_0, g_1$. Then, after averaging over $n^A$, these 3 diagrams give rise to
\begin{equation}
 2{\rm (a)} +  2{\rm (b)} +  2{\rm (c)} = \frac{1}{\lambda}\delta^{AB} q_{12}\,.
\end{equation}
The comparison with \eqref{2ptexpa} indicates that
\begin{equation}
\label{q12}
 q_{12}=c_2 - \frac{1}{3}d_2 {\rm L}_{12} +\frac{d_1^2}{6} {\rm L}_{12}^2\,.
\end{equation}
In order to know the precise values of $d_2$ and $c_2$ one would need to compute the 1-loop corrections mentioned before. However, as we will see in the next section, these precise values would not be necessary for testing the conformal covariance of the 4-point function. For the coefficient in  front of ${\rm L}_{12}^2$ to be the one indicated in \eqref{q12}, 
\begin{equation}
    f_2 = \frac{1}{4\pi(1-{\cal B}^2)^2}\,,
\label{expectedf2}
\end{equation} 
is needed.

\subsection{4-point functions}

As stated in the Introduction,  the primary goal of our article is to ascertain whether the dual theory on the line, defined by fluctuations in AdS$_2$ with mixed boundary conditions \eqref{eq:quadraticproblem}, exhibits conformal behaviour. Expressing the 4-point function of embedding coordinates as
\begin{align}
 \langle Y^A\left(t_1\right) Y^B\left(t_2\right) Y^C\left(t_3\right) Y^D\left(t_4\right)\rangle &=
  \frac{C_Y^2\, G^{ABCD}}{(t_2-t_1)^{2\Delta}(t_4-t_3)^{2\Delta}}\,,
  \end{align}
conformal symmetry would require that
\begin{equation}
G^{ABCD} = G^{ABCD}(\lambda,u)\,,     
\end{equation}
where $u$ is the unique independent cross-ratio
\begin{equation}
    u = \frac{(t_1-t_2)(t_3-t_4)}{(t_1-t_3)(t_2-t_4)}\,.
\end{equation}
Following \cite{Beccaria:2019dws} we can split $G^{ABCD}$ into singlet, symmetric traceless and antisymmetric tensors
\begin{align}
        G^{ABCD} =& \ G_{\sf S}\delta^{AB}\delta^{CD}+G_{\sf T}(\delta^{AC}\delta^{BD}+\delta^{AD}\delta^{BC}-\tfrac{2}{3}\delta^{AB}\delta^{CD})
     \nonumber
     \\  & +G_{\sf A}(\delta^{AC}\delta^{BD}-\delta^{AD}\delta^{BC} )\,,
\end{align}
and concentrate on the singlet, as the other tensors can be related to $G_{\sf S}$ using crossing transformations. For $SO(3)$, which is the case of interest to us, one has \cite{Beccaria:2019dws}
\begin{align}
G_{\sf T}(u) & =-\frac{3}{10} G_{\sf S}(u)
+\frac{9}{20}\left(u^{2\Delta} G_{\sf S}(\tfrac{1}{1-u})+
(\tfrac{u}{1-u})^{2\Delta} G_{\sf S}(\tfrac{u}{1-u})\right)\,,
\\
G_{\sf A} (u) & = \frac{3}{4}\left(u^{2\Delta} G_{\sf S}(\tfrac{1}{1-u})-
(\tfrac{u}{1-u})^{2\Delta} G_{\sf S}(\tfrac{u}{1-u})\right)\,.
\end{align}

To compute the singlet, we just need to consider
\begin{align}
\label{YYYYAABB}
    \langle Y^A\left(t_1\right) Y^A\left(t_2\right) Y^B\left(t_3\right) Y^B\left(t_4\right)\rangle &=
  \frac{9 C_Y^2 G_{\sf S}(\lambda,u)}{(t_2-t_1)^{2\Delta}(t_4-t_3)^{2\Delta}} \nonumber
  \\
  &= 1 + \frac{Q^{(1)}}{\sqrt\lambda}+ 
  \frac{Q^{(2)}}{(\sqrt\lambda)^2} +
   \frac{Q^{(3)}}{(\sqrt\lambda)^3}+
   {\cal O}(\tfrac{1}{\lambda^2})\,.
\end{align}
We can expand 
\begin{equation}
    G_{\sf S} =  1 + \frac{G_{\sf S}^{(1)}}{\sqrt\lambda}+ 
  \frac{G_{\sf S}^{(2)}}{(\sqrt\lambda)^2} +
   \frac{G_{\sf S}^{(3)}}{(\sqrt\lambda)^3}+
   {\cal O}(\tfrac{1}{\lambda^2})\,,
   \label{GSexpansion}
\end{equation}
and, if each successive order is a function of the cross-ratio $u$, this can be regarded as an indication of the conformal symmetry of the theory on the line. To obtain the successive
$G_{\sf S}^{(i)}$ one needs to compute the successive $Q^{(i)}$, and for this one has to use once again the expansion
\begin{eqnarray}
     Y^A= n^A + \frac{\sqrt{2\pi}}{\lambda^{1/4}} y^A -\frac{\pi}{\sqrt{\lambda}} y^2 n^A + {\cal O}(\tfrac{1}{\lambda})\,.
     \label{Yexpa2}
\end{eqnarray}

\subsubsection*{Order $1/\sqrt{\lambda}$}
 Replacing \eqref{DeltaCexpansions} and \eqref{GSexpansion} in \eqref{YYYYAABB}, and keeping the order $1/\sqrt{\lambda}$, we obtain
\begin{equation}
  G_{\sf S}^{(1)}  = Q^{(1)} + d_1\left({\rm L}_{12}+{\rm L}_{34}\right) \,. 
\end{equation}
The diagrams contributing to $Q^{(1)}$ come from 
$\langle y^A(t_1) y^A(t_2)n^Bn^B\rangle_0$ and $\langle n^A n^A y^B(t_3) y^B(t_4)\rangle_0$, depicted in Fig. \ref{fig:Q1}
\begin{figure}[h!]
    \centering
    \begin{tikzpicture}[scale=0.8]
    \draw[dashed,thick] (0,0) circle [radius=0.8cm];
     \draw[thick] (0.8*0.707,0.8*0.707)
     --(-0.8*0.707,0.8*0.707);
    \draw[fill=black] (0.8*0.707,0.8*0.707) circle[radius=0.05cm];
     \node[right] at (0.8*0.707,0.8*0.707) {$y^A$};
    \draw[fill=white] (0.8*0.707,-0.8*0.707) circle[radius=0.05cm];
    \node[right] at (0.8*0.707,-0.8*0.707) {$n^B$};
    \draw[fill=black] (-0.8*0.707,0.8*0.707) circle[radius=0.05cm];
    \node[left] at (-0.8*0.707,0.8*0.707) {$y^A$};
    \draw[fill=white] (-0.8*0.707,-0.8*0.707) circle[radius=0.05cm];
    \node[left] at 
    (-0.8*0.707,-0.8*0.707)
    {$n^B$};
    \node at (0,-1.15) {(a)};
    \end{tikzpicture}
    \hspace{0.6cm}
    %%%%%%%%
    \begin{tikzpicture}[scale=0.8]
    \draw[dashed,thick] (0,0) circle [radius=0.8cm];
     \draw[thick] (0.8*0.707,-0.8*0.707)
     --(-0.8*0.707,-0.8*0.707);
    \draw[fill=black] (0.8*0.707,-0.8*0.707) circle[radius=0.05cm];
     \node[right] at (0.8*0.707,-0.8*0.707) {$y^B$};
    \draw[fill=white] (0.8*0.707,0.8*0.707) circle[radius=0.05cm];
    \node[right] at (0.8*0.707,0.8*0.707) {$n^A$};
    \draw[fill=black] (-0.8*0.707,-0.8*0.707) circle[radius=0.05cm];
    \node[left] at (-0.8*0.707,-0.8*0.707) {$y^B$};
    \draw[fill=white] (-0.8*0.707,0.8*0.707) circle[radius=0.05cm];
    \node[left] at 
    (-0.8*0.707,0.8*0.707)
    {$n^A$};
    \node at (0,-1.15) {(b)};
    \end{tikzpicture}
    \caption{Diagrams contributing to $Q^{(1)}$.}
    \label{fig:Q1}
\end{figure}
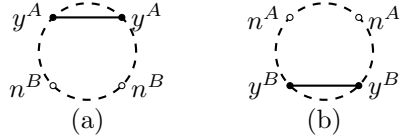
\begin{equation}
 3{\rm(a)}=-\frac{1}{\sqrt\lambda}\frac{2{\rm L}_{12}}{1-{\cal B}^2} \,,\qquad
  3{\rm(b)}=-\frac{1}{\sqrt\lambda}\frac{2{\rm L}_{34}}{1-{\cal B}^2} \,,
\end{equation}
Other diagrams with just one propagator are vanishing. For example
\begin{equation}
  \langle n^A y^A(t_2) y^B(t_3) n^B\rangle =0\,.
\end{equation}
because $n^A y^A = 0$. As a consequence,
\begin{equation}
    Q^{(1)}= - d_1\left({\rm L}_{12}+{\rm L}_{34}\right) \quad \Rightarrow \quad G_{\sf S}^{(1)} = 0\,. 
\end{equation}

\subsubsection*{Order $1/{\lambda}$}
Proceeding with the order $1/{\lambda}$, we obtain
\begin{align}
   G_{\sf S}^{(2)} &= Q^{(2)} - 6 c_2 + d_2\left({\rm L}_{12}+{\rm L}_{34}\right)-\frac{d_1^2}{2}\left({\rm L}_{12}+{\rm L}_{34}\right)^2 \nonumber\\&=
   Q^{(2)} - 3 q_{12} - 3q_{34} - d_1^2 {\rm L}_{12}{\rm L}_{34}\,,
   \label{Q2G2}
\end{align}
where we have used the notation $q_{ij}$ introduced in \eqref{q12}. 

Let us then analyze all the diagrams contributing to $Q^{(2)}$. We can organize them according to the number of insertion points containing just a vector $n^A$, represented with a white dot in the diagrams. With $n^A$ inserted in 2 out of the 4 points we have diagrams with either a double propagator or 1 propagator
with 1-loop corrections. 
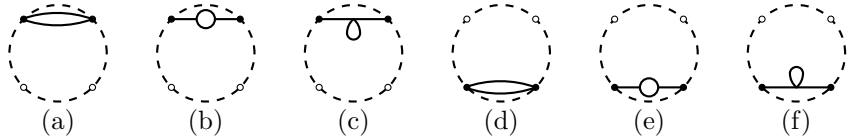
\begin{figure}[h!]
    \centering
    \begin{tikzpicture}[scale=0.8]
    \draw[dashed,thick] (0,0) circle [radius=0.8cm];
    \draw[thick,cap=round] (0.8*0.707,0.8*0.707) arc (70:110:1.7cm);
    \draw[thick,cap=round] (0.8*0.707,0.8*0.707) arc (-70:-110:1.7cm);
    \draw[fill=black] (0.8*0.707,0.8*0.707) circle[radius=0.05cm];
    \draw[fill=white] (0.8*0.707,-0.8*0.707) circle[radius=0.05cm];
    \draw[fill=black] (-0.8*0.707,0.8*0.707) circle[radius=0.05cm];
    \draw[fill=white] (-0.8*0.707,-0.8*0.707) circle[radius=0.05cm];
    \node at (0,-1.15) {(a)};
    \end{tikzpicture}
    \hspace{0.4cm}
    %%%%%%%%%%%%%
    \begin{tikzpicture}[scale=0.8]
    \draw[dashed,thick] (0,0) circle [radius=0.8cm];
     \draw[thick] (0.8*0.707,0.8*0.707)
     --(-0.8*0.707,0.8*0.707);
\draw[thick,fill=white] (0,0.8*0.707)
     circle[radius=0.15cm];
       \draw[fill=black] (0.8*0.707,0.8*0.707) circle[radius=0.05cm];
     \draw[fill=white] (0.8*0.707,-0.8*0.707) circle[radius=0.05cm];
    \draw[fill=black] (-0.8*0.707,0.8*0.707) circle[radius=0.05cm];
    \draw[fill=white] (-0.8*0.707,-0.8*0.707) circle[radius=0.05cm];
    \node at (0,-1.15) {(b)};
    \end{tikzpicture}
    \hspace{0.4cm}
    %%%%%%%%%%%%%
    \begin{tikzpicture}[scale=0.8]
    \draw[dashed,thick] (0,0) circle [radius=0.8cm];
     \draw[thick] (0.8*0.707,0.8*0.707)
     --(-0.8*0.707,0.8*0.707);
\draw[thick,cap=round] (0,0.8*0.707) arc (50:-0:0.3cm);
\draw[thick,cap=round] (0,0.8*0.707) arc (-230:-180:0.3cm);
\draw[thick,cap=round] (0.108,0.335) arc (0:-180:0.108cm);
    \draw[fill=black] (0.8*0.707,0.8*0.707) circle[radius=0.05cm];
    \draw[fill=white] (0.8*0.707,-0.8*0.707) circle[radius=0.05cm];
    \draw[fill=black] (-0.8*0.707,0.8*0.707) circle[radius=0.05cm];
    \draw[fill=white] (-0.8*0.707,-0.8*0.707) circle[radius=0.05cm];
    \node at (0,-1.15) {(c)};
    \end{tikzpicture}
    \hspace{0.4cm}
    %%%%%%%%
    \begin{tikzpicture}[scale=0.8]
    \draw[dashed,thick] (0,0) circle [radius=0.8cm];
    \draw[thick,cap=round] (0.8*0.707,-0.8*0.707) arc (70:110:1.7cm);
    \draw[thick,cap=round] (0.8*0.707,-0.8*0.707) arc (-70:-110:1.7cm);
    \draw[fill=black] (0.8*0.707,-0.8*0.707) circle[radius=0.05cm];
    \draw[fill=white] (0.8*0.707,0.8*0.707) circle[radius=0.05cm];
    \draw[fill=black] (-0.8*0.707,-0.8*0.707) circle[radius=0.05cm];
    \draw[fill=white] (-0.8*0.707,0.8*0.707) circle[radius=0.05cm];
    \node at (0,-1.15) {(d)};
    \end{tikzpicture}
    \hspace{0.4cm}
        %%%%%%%%%%%%%
    \begin{tikzpicture}[scale=0.8]
    \draw[dashed,thick] (0,0) circle [radius=0.8cm];
     \draw[thick] (0.8*0.707,-0.8*0.707)
     --(-0.8*0.707,-0.8*0.707);
     \draw[thick,fill=white] (0,-0.8*0.707)
     circle[radius=0.15cm];
    \draw[fill=white] (0.8*0.707,0.8*0.707) circle[radius=0.05cm];
    \draw[fill=black] (0.8*0.707,-0.8*0.707) circle[radius=0.05cm];
    \draw[fill=white] (-0.8*0.707,0.8*0.707) circle[radius=0.05cm];
    \draw[fill=black] (-0.8*0.707,-0.8*0.707) circle[radius=0.05cm];
    \node at (0,-1.15) {(e)};
    \end{tikzpicture}
    \hspace{0.4cm}
    %%%%%%%%%%%%%
    \begin{tikzpicture}[scale=0.8]
    \draw[dashed,thick] (0,0) circle [radius=0.8cm];
     \draw[thick] (0.8*0.707,-0.8*0.707)
     --(-0.8*0.707,-0.8*0.707);
\draw[thick,cap=round] (0,-0.8*0.707) arc (-50:-0:0.3cm);
    \draw[thick,cap=round] (0,-0.8*0.707) arc (230:180:0.3cm);
    \draw[thick,cap=round] (0.108,-0.335) arc (0:180:0.108cm);
    \draw[fill=black] (0.8*0.707,-0.8*0.707) circle[radius=0.05cm];
    \draw[fill=white] (0.8*0.707,0.8*0.707) circle[radius=0.05cm];
    \draw[fill=black] (-0.8*0.707,-0.8*0.707) circle[radius=0.05cm];
    \draw[fill=white] (-0.8*0.707,0.8*0.707) circle[radius=0.05cm];
    \node at (0,-1.15) {(f)};
    \end{tikzpicture}
       \caption{Some diagrams contributing to $Q^{(2)}_2$.}
    \label{fig:enter-label-Q22a0}
\end{figure}
The diagrams in Fig. \ref{fig:enter-label-Q22a0} are entirely analogous to those computing the order $1/\lambda$ in the 2-point function. Thus
\begin{equation}
   4{\rm (a)}+4{\rm (b)}+4{\rm (c)}+
   4{\rm (d)}+4{\rm (e)}+4{\rm (f)} = \frac{3}{\lambda}(q_{12}+q_{34})\,.
\end{equation}
\begin{figure}[h!]
    \centering
     \begin{tikzpicture}[scale=0.8]
    \draw[dashed,thick] (0,0) circle [radius=0.8cm];
    \draw[thick,cap=round] (0.8*0.707,0.8*0.707) arc (160:200:1.7cm);
    \draw[thick,cap=round] (0.8*0.707,0.8*0.707) arc (20:-20:1.7cm);
    \draw[fill=black] (0.8*0.707,0.8*0.707) circle[radius=0.05cm];
     \draw[fill=black] (0.8*0.707,-0.8*0.707) circle[radius=0.05cm];
    \draw[fill=white] (-0.8*0.707,0.8*0.707) circle[radius=0.05cm];
    \draw[fill=white] (-0.8*0.707,-0.8*0.707) circle[radius=0.05cm];
    \node at (0,-1.15) {(a)};
    \end{tikzpicture}
    \hspace{0.4cm}
    %%%%%%%%%
    \begin{tikzpicture}[scale=0.8]
    \draw[dashed,thick] (0,0) circle [radius=0.8cm];
    \draw[thick,cap=round] (-0.8*0.707,0.8*0.707) arc (160:200:1.7cm);
    \draw[thick,cap=round] (-0.8*0.707,0.8*0.707) arc (20:-20:1.7cm);
    \draw[fill=black] (-0.8*0.707,-0.8*0.707) circle[radius=0.05cm];
    \draw[fill=black] (-0.8*0.707,0.8*0.707) circle[radius=0.05cm];
    \draw[fill=white] (0.8*0.707,-0.8*0.707) circle[radius=0.05cm];
    \draw[fill=white] (0.8*0.707,0.8*0.707) circle[radius=0.05cm];
    \node at (0,-1.15) {(b)};
    \end{tikzpicture}
%%%%%%%%%
\hspace{0.4cm}
\begin{tikzpicture}[scale=0.8]
    \draw[dashed,thick] (0,0) circle [radius=0.8cm];
    \draw[thick,cap=round] (-0.8*0.707,0.8*0.707) arc (205:245:2.32cm);
    \draw[thick,cap=round] (-0.8*0.707,0.8*0.707) arc (65:25:2.32cm);
    \draw[fill=white] (-0.8*0.707,-0.8*0.707) circle[radius=0.05cm];
    \draw[fill=black] (-0.8*0.707,0.8*0.707) circle[radius=0.05cm];
    \draw[fill=black] (0.8*0.707,-0.8*0.707) circle[radius=0.05cm];
    \draw[fill=white] (0.8*0.707,0.8*0.707) circle[radius=0.05cm];
    \node at (0,-1.15) {(c)};
    \end{tikzpicture}
    %%%%%%
    \hspace{0.4cm}
    \begin{tikzpicture}[scale=0.8]
    \draw[dashed,thick] (0,0) circle [radius=0.8cm];
    \draw[thick,cap=round] (0.8*0.707,0.8*0.707) arc (115:155:2.32cm);
    \draw[thick,cap=round] (0.8*0.707,0.8*0.707) arc (-25:-65:2.32cm);
    \draw[fill=black] (-0.8*0.707,-0.8*0.707) circle[radius=0.05cm];
    \draw[fill=white] (-0.8*0.707,0.8*0.707) circle[radius=0.05cm];
    \draw[fill=white] (0.8*0.707,-0.8*0.707) circle[radius=0.05cm];
    \draw[fill=black] (0.8*0.707,0.8*0.707) circle[radius=0.05cm];
    \node at (0,-1.15) {(d)};
    \end{tikzpicture}
    %%%%%%%%%
       \caption{Some diagrams contributing to $Q^{(2)}_2$.}
    \label{fig:enter-label-Q22b0}
\end{figure}
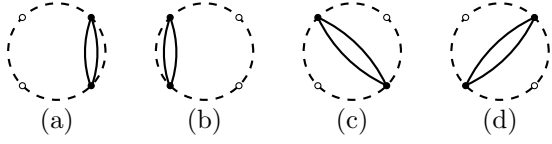
%%%%%%%%%%%%%%%%%%%%%%%%%%%%%%%%%%%%%%%%%%

For diagrams of this sort, when contracting other pairs than 12 and 34, only the ones with the double propagator - represented in Fig. \ref{fig:enter-label-Q22b0} - are non-vanishing. Putting all the diagrams in Fig. \ref{fig:enter-label-Q22a0} and Fig. \ref{fig:enter-label-Q22b0} together we obtain
\begin{align}
Q_{2}^{(2)} =  3 \left( q_{13} + q_{34} \right) 
+ \frac{1}{(1-{\cal B}^2)^2} \left({\rm L}^2_{14} + {\rm L}^2_{13} + {\rm L}^2_{23} + {\rm L}^2_{24} + 4 \pi^2 {\cal B}^2  \right) \,.
\end{align}

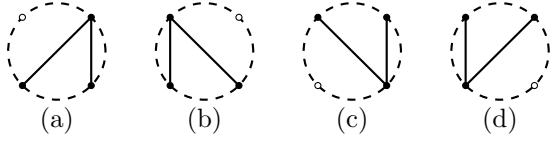
\begin{figure}[h!]
    \centering
    \begin{tikzpicture}[scale=0.8]
    \draw[dashed,thick] (0,0) circle [radius=0.8cm];
    \draw[thick] (0.8*0.707,0.8*0.707)
     --(0.8*0.707,-0.8*0.707);
     \draw[thick] (0.8*0.707,0.8*0.707)
     --(-0.8*0.707,-0.8*0.707);
    \draw[fill=black] (0.8*0.707,0.8*0.707) circle[radius=0.05cm];
   \draw[fill=black] (0.8*0.707,-0.8*0.707) circle[radius=0.05cm];
\draw[fill=white] (-0.8*0.707,0.8*0.707) circle[radius=0.05cm];
   \draw[fill=black] (-0.8*0.707,-0.8*0.707) circle[radius=0.05cm];
    \node at (0,-1.15) {(a)};
    \end{tikzpicture}
    %%%%%%%%%
    \hspace{0.4cm}
    \begin{tikzpicture}[scale=0.8]
    \draw[dashed,thick] (0,0) circle [radius=0.8cm];
    \draw[thick] (-0.8*0.707,0.8*0.707)
     --(0.8*0.707,-0.8*0.707);
     \draw[thick] (-0.8*0.707,0.8*0.707)
     --(-0.8*0.707,-0.8*0.707);
    \draw[fill=white] (0.8*0.707,0.8*0.707) circle[radius=0.05cm];
   \draw[fill=black] (0.8*0.707,-0.8*0.707) circle[radius=0.05cm];
   \draw[fill=black] (-0.8*0.707,0.8*0.707) circle[radius=0.05cm];
    \draw[fill=black] (-0.8*0.707,-0.8*0.707) circle[radius=0.05cm];
    \node at (0,-1.15) {(b)};
    \end{tikzpicture}
%%%%%%%%%
\hspace{0.4cm}
    \begin{tikzpicture}[scale=0.8]
    \draw[dashed,thick] (0,0) circle [radius=0.8cm];
    \draw[thick] (0.8*0.707,-0.8*0.707)
     --(0.8*0.707,0.8*0.707);
     \draw[thick] (0.8*0.707,-0.8*0.707)
     --(-0.8*0.707,0.8*0.707);
    \draw[fill=black] (0.8*0.707,-0.8*0.707) circle[radius=0.05cm];
    \draw[fill=black] (0.8*0.707,0.8*0.707) circle[radius=0.05cm];
    \draw[fill=white] (-0.8*0.707,-0.8*0.707) circle[radius=0.05cm];
    \draw[fill=black] (-0.8*0.707,0.8*0.707) circle[radius=0.05cm];
    \node at (0,-1.15) {(c)};
    \end{tikzpicture}
    %%%%%%%%%
    \hspace{0.4cm}
    \begin{tikzpicture}[scale=0.8]
    \draw[dashed,thick] (0,0) circle [radius=0.8cm];
    \draw[thick] (-0.8*0.707,-0.8*0.707)
     --(0.8*0.707,0.8*0.707);
     \draw[thick] (-0.8*0.707,-0.8*0.707)
     --(-0.8*0.707,0.8*0.707);
    \draw[fill=white] (0.8*0.707,-0.8*0.707) circle[radius=0.05cm];
    \draw[fill=black] (0.8*0.707,0.8*0.707) circle[radius=0.05cm];
    \draw[fill=black] (-0.8*0.707,0.8*0.707) circle[radius=0.05cm];
    \draw[fill=black] (-0.8*0.707,-0.8*0.707) circle[radius=0.05cm];
    \node at (0,-1.15) {(d)};
    \end{tikzpicture}
    %%%%%%%%%
       \caption{Diagrams contributing to $Q^{(2)}_1$.}
    \label{fig:enter-label-Q21}
\end{figure}

Non-vanishing diagrams with just one insertion of the type $n^A$ are shown in Fig. \ref{fig:enter-label-Q21}. Their total contribution is
\begin{align}
Q_{1}^{(2)} = -\frac{2}{(1-{\cal B}^2)^2} & \left({\rm L}_{14}{\rm L}_{13}  + {\rm L}_{23}{\rm L}_{24} + {\rm L}_{14} {\rm L}_{23}  +  {\rm L}_{13}{\rm L}_{24} \right) \nonumber
\\
-\frac{2\pi^2 {\cal B}^2}{(1-{\cal B}^2)^2} & \left( {\rm S}_{14} {\rm S}_{13} + {\rm S}_{23} {\rm S}_{24}  +  {\rm S}_{14} {\rm S}_{23}  +  {\rm S}_{13} {\rm S}_{24}  \right)\,,
\end{align}
where
\begin{equation}
  {\rm S}_{ij}:= {\rm sign}(t_i-t_j)\,.  
\end{equation}

%%%%%%%%%%%%%%%%%%%%%%%%%%%%%%%%

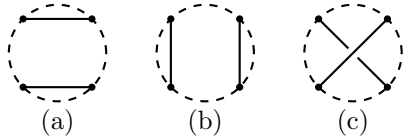
\begin{figure}[h!]
    \centering
    \begin{tikzpicture}[scale=0.8]
    \draw[dashed,thick] (0,0) circle [radius=0.8cm];
     \draw[thick] (0.8*0.707,0.8*0.707)
     --(-0.8*0.707,0.8*0.707);
    \draw[thick] (0.8*0.707,-0.8*0.707)
     --(-0.8*0.707,-0.8*0.707);
    \draw[fill=black] (0.8*0.707,0.8*0.707) circle[radius=0.05cm];
    \draw[fill=black] (0.8*0.707,-0.8*0.707) circle[radius=0.05cm];
    \draw[fill=black] (-0.8*0.707,0.8*0.707) circle[radius=0.05cm];
    \draw[fill=black] (-0.8*0.707,-0.8*0.707) circle[radius=0.05cm];
    \node at (0,-1.15) {(a)};
    \end{tikzpicture}
    %%%%%%%%
    \hspace{0.4cm}
    \begin{tikzpicture}[scale=0.8]
    \draw[dashed,thick] (0,0) circle [radius=0.8cm];
     \draw[thick] (0.8*0.707,0.8*0.707)
     --(0.8*0.707,-0.8*0.707);
     \draw[thick] (-0.8*0.707, 0.8*0.707)
     --(-0.8*0.707,-0.8*0.707);
    \draw[fill=black] (0.8*0.707,-0.8*0.707) circle[radius=0.05cm];
    \draw[fill=black] (0.8*0.707,0.8*0.707) circle[radius=0.05cm];
    \draw[fill=black] (-0.8*0.707,-0.8*0.707) circle[radius=0.05cm];
    \draw[fill=black] (-0.8*0.707,0.8*0.707) circle[radius=0.05cm];
    \node at (0,-1.15) {(b)};
    \end{tikzpicture}
\hspace{0.4cm}
 \begin{tikzpicture}[scale=0.8]
    \draw[dashed,thick] (0,0) circle [radius=0.8cm];
     \draw[thick] (-0.8*0.707,0.8*0.707)
     --(0.8*0.707,-0.8*0.707);
     \draw[white,fill=white] (0,0) circle [radius=0.1cm];
     \draw[thick] (-0.8*0.707, -0.8*0.707)
     --(0.8*0.707,0.8*0.707);
    \draw[fill=black] (0.8*0.707,-0.8*0.707) circle[radius=0.05cm];
    \draw[fill=black] (0.8*0.707,0.8*0.707) circle[radius=0.05cm];
    \draw[fill=black] (-0.8*0.707,-0.8*0.707) circle[radius=0.05cm];
    \draw[fill=black] (-0.8*0.707,0.8*0.707) circle[radius=0.05cm];
    \node at (0,-1.15) {(c)};
    \end{tikzpicture}
    \caption{Diagrams contributing to $Q^{(2)}_0$.}
    \label{fig:enter-label-Q20}
 \end{figure}

The last contribution to this order comes from diagrams with $y^A$ in the 4 insertion points, as represented in Fig. \ref{fig:enter-label-Q20}. They give 
\begin{align} 
Q_{0}^{(2)} =  \frac{1}{(1-{\cal B}^2)^2} \left( 4 {\rm L}_{12} {\rm L}_{34} + 2  {\rm L}_{14}{\rm L}_{23} + 2 \pi^2  {\cal B}^2 {\rm S}_{14} {\rm S}_{23} + 2  {\rm L}_{13}{\rm L}_{24} + 2\pi^2 {\cal B}^2 {\rm S}_{13} {\rm S}_{24} \right)\,. 
\end{align}

Now, we can combine all the contributions to $Q^{(2)}$. Note that the terms proportional to $q_{12}$, $q_{34}$ and $ L_{12} L_{34}$ in the rhs of \eqref{Q2G2} cancel with some of the contributions to $Q^{(2)}$. As a result, we obtain the following expression for $G_{\sf S}^{(2)}$
 \begin{align} 
   G_{\sf S}^{(2)} &= \frac{1}{(1-{\cal B}^2)^2} \left({\rm L}_{14} + {\rm L}_{23} - {\rm L}_{13} - {\rm L}_{24}\right)^2 + \frac{\pi^2{\cal B}^2}{(1-{\cal B}^2)^2} \left({\rm S}_{14} + {\rm S}_{23} - {\rm S}_{13} - {\rm S}_{24}\right)^2\,,
   \label{G2bis}
 \end{align}
which can be re-expressed as a function of the cross-ratio 
 \begin{align}
   G_{\sf S}^{ (2)}(u) = \frac{4}{(1-{\cal B}^2)^2}  \left[\log^2\left( {1 - u}\right) + \pi^2  {\cal B}^2  \Theta(u-1) \right]\,,
 \end{align}
where $\Theta$ is the Heaviside step function. The fact that the first non-trivial contribution to $G_{\sf S}$ is indeed just a function of the cross-ratio serves as evidence for the conformal covariance of the 4-point function. Another interesting observation about this result is that it verifies
\begin{align}
    G_{\sf S}^{ (2)}(u) = G_{\sf S}^{ (2)}(\tfrac{u}{u-1})\,, 
\end{align}
a crossing symmetry relation from the exchange $t_3\longleftrightarrow t_4$.

It is worth noting that, at this order, all contributions arise from reducible diagrams originated from quadratic terms in the action of fluctuations. A more robust confirmation of conformal covariance would be obtained exploring the next order, which incorporates contributions from connected Witten diagrams involving quartic terms.

\subsubsection*{Order $1/(\sqrt{\lambda})^3$}
Expanding eq. \eqref{YYYYAABB} to order
$\frac{1}{\lambda^{3/2}}$, we obtain
\begin{align}
\!\!G_{\sf S}^{(3)} = \ & Q^{(3)} - 6 c_3
+ \left(d_3 + 6 d_1 c_2 + d_1 G_{\sf S}^{(2)}\right)\left({\rm L}_{12}+{\rm L}_{34}\right)-{d_1d_2}\left({\rm L}_{12}+{\rm L}_{34}\right)^2+
\frac{d_1^3}{6}\left({\rm L}_{12}+{\rm L}_{34}\right)^3\!\!.
\label{GS3}
\end{align}

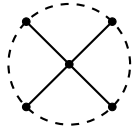
\begin{figure}[h!]
    \centering
 \begin{tikzpicture}[scale=1]
    \draw[dashed,thick] (0,0) circle [radius=0.8cm];
     \draw[thick] (-0.8*0.707,0.8*0.707)
     --(0.8*0.707,-0.8*0.707);
     \draw[fill=black] (0,0) circle [radius=0.05cm];
     \draw[thick] (-0.8*0.707, -0.8*0.707)
     --(0.8*0.707,0.8*0.707);
    \draw[fill=black] (0.8*0.707,-0.8*0.707) circle[radius=0.05cm];
    \draw[fill=black] (0.8*0.707,0.8*0.707) circle[radius=0.05cm];
    \draw[fill=black] (-0.8*0.707,-0.8*0.707) circle[radius=0.05cm];
    \draw[fill=black] (-0.8*0.707,0.8*0.707) circle[radius=0.05cm];
    \end{tikzpicture}
    \caption{Irreducible diagrams contributing to $Q^{(3)}$.}
    \label{fig:enter-label-Q3}
 \end{figure}

At this order we will have to include irreducible diagrams from the quartic terms of the action contributing to $Q^{(3)}$, 
like the one depicted in Fig. \ref{fig:enter-label-Q3}. There are 4 different types of quartic vertices
\begin{alignat}{3} \label{vertices1}
 V_1 = & \frac{\pi}{\sqrt \lambda} y^A y^B \partial_\mu y^A
    \partial^\mu y^B\,, \qquad
 && V_2 =       \frac{\pi}{4\sqrt \lambda}(\partial_\mu y^A
    \partial^\mu y^A)^2\,,
 \\
  V_3 = &  -\frac{\pi}{2\sqrt \lambda}\partial_\mu y^A
    \partial^\mu y^B \partial_\nu y^A
    \partial^\nu y^B\,,\qquad
 && V_{4} =  - \frac{\pi}{\sqrt{\lambda}}\frac{1}{\sqrt{g}}\mathcal{B} \epsilon^{ABC}  n^A y^2 \partial_z y^B \partial_t y^C\,.
 \label{vertices2}
\end{alignat}

All these diagrams involve a spacetime integration of products of bulk-to-boundary propagators. The diagram with $V_1$ will be more intricate, as two propagators, appearing without derivatives, introduce logarithmic or arc tangent factors. To avoid this complication, we can compute 
$\partial_{t_1} \partial_{t_2} \partial_{t_3} \partial_{t_4} \left \langle Y^{A}(t_1) Y^{A}(t_2) Y^{B}(t_3) Y^{B}(t_4) \right \rangle$. Therefore, instead of \eqref{GS3}, we will consider
\begin{align}
\partial_{t_1} \partial_{t_2} \partial_{t_3} \partial_{t_4} G_{\sf S}^{(3)} = \ & \partial_{t_1} \partial_{t_2} \partial_{t_3} \partial_{t_4} Q^{(3)} 
+  d_1 \partial_{t_1} \partial_{t_2} \partial_{t_3} \partial_{t_4}\left( G_{\sf S}^{(2)}\left({\rm L}_{12}+{\rm L}_{34}\right)\right)
\nonumber 
\\
& -2 {d_1d_2} \partial_{t_1} \partial_{t_2} \partial_{t_3} \partial_{t_4}\left({\rm L}_{12}{\rm L}_{34}\right)
+ \frac{d_1^3}{2}
\partial_{t_1} \partial_{t_2} \partial_{t_3} \partial_{t_4}\left(
{\rm L}_{12}^2{\rm L}_{34}+{\rm L}_{12}{\rm L}_{34}^2\right)\,.
\label{ddddGS3}
\end{align}

It is straightforward to check that,
acting with the four derivatives on a function of the cross-ratio, we obtain\footnote{Where
\begin{equation}
    \tilde F(u) =
    u^2\left[u^2(u-1)^2F^{(4)}(u)+4u(u-1)(2u-1)F^{(3)}(u)+(2+14u(u-1))F''(u)
    +2u^2(2u-1)F'(u)\right]\,.
    \nonumber
\end{equation}
} 
\begin{equation}
   \partial_{t_1} \partial_{t_2} \partial_{t_3} \partial_{t_4} F(u) =
   \frac{\tilde F(u)}{(t_1-t_2)^2(t_3-t_4)^2}\,.
\end{equation}
Therefore, we have to check whether the rhs of \eqref{ddddGS3} is of this form. 

Let us begin by computing the irreducible contact diagrams.  For the  vertex $V_1$ we have to compute 
\begin{align}
   \partial_{t_1} \partial_{t_2} \partial_{t_3} \partial_{t_4} Q^{(3)}_{irred,1} = & \ 
 - {4\pi^3}\int d^2\sigma\sqrt{g}
   \langle
   \partial_{t_1}y^A \partial_{t_2}y^A
   \partial_{t_3}y^B \partial_{t_4}y^B
   y^C y^D\partial_\mu y^C\partial^\mu y^D
   \rangle_0\,.
\end{align}
This can be explicitly computed and the result is
\begin{align}
   t_{12}^2t_{34}^2 \partial_{t_1} \partial_{t_2} & \partial_{t_3} \partial_{t_4} Q^{(3)}_{irred,1} =  
   -\frac{8 u^2\left[u(u-2)+2\right]\log(u)}{(1-{\cal B}^2)^2(1-u)^2} +
   \frac{8 u^2\log(1-u)}{(1-{\cal B}^2)^2}
   -\frac{8 u^2}{(1-{\cal B}^2)^2(1-u)}
   \nonumber\\
   &-\frac{32 t_{12}^2t_{34}^2}{(1-{\cal B}^2)^3}\left(\frac{1}{t_{12}t_{23}^2t_{24}}-\frac{1}{t_{12}t_{13}^2t_{14}}\right)
   -
   \frac{16(3-{\cal B}^2)}{(1-{\cal B}^2)^4}t_{12}^2t_{34}^2\left(\frac{1}{t_{12}t_{23}t_{24}^2}-\frac{1}{t_{12}t_{13}t_{14}^2}\right)
   \nonumber
   \\
   &
   -16\frac{(1+{\cal B}^2)}{(1-{\cal B}^2)^4}t_{12}^2t_{34}^2\left(\frac{1}{t_{13}t_{23}^2t_{34}} +\frac{1}{t_{13}^2t_{23}t_{34}} -\frac{1}{t_{12}^2t_{23}t_{24}}
    -\frac{1}{t_{12}^2t_{13}t_{14}}-\frac{2}{t_{13}t_{23}t_{34}^2}\right).
   \label{Q3V1}
\end{align}
One comment is in order about this result. As we have assumed that all $t_i$ are different to perform the integrals, we have lost track of contact-like terms in the form of Dirac deltas or derivatives of them. We know this type of terms appear in the reducible diagrams, from derivatives acting on the sign functions present in the boundary-to-boundary propagators.

In relation with our motivation for the study of this 4-point function, we should notice the appearance of terms that are not functions of the cross-ratio. Similar anomalous contributions appear in the vertex originated from the coupling with the Kalb-Ramond field
\begin{align}
  \partial_{t_1} \partial_{t_2} \partial_{t_3} \partial_{t_4} Q^{(3)}_{irred,4} = & \ 
   4\pi^3
   {\cal B}n^C\epsilon^{CDE}
   \int d^2\sigma
   \langle
   \partial_{t_1}y^A \partial_{t_2}y^A
   \partial_{t_3}y^B \partial_{t_4}y^B
   y^F y^F \partial_{z}y^D \partial_{t}y^E
   \rangle_0   \,.
   \label{Q3V4}
\end{align}
The remaining irreducible contributions are not anomalous, {\it i.e.} $ t_{12}^2t_{34}^2 \partial_{t_1} \partial_{t_2} \partial_{t_3} \partial_{t_4} Q^{(3)}_{irred,2}$ and $ t_{12}^2t_{34}^2 \partial_{t_1} \partial_{t_2} \partial_{t_3} \partial_{t_4} Q^{(3)}_{irred,3}$ are just functions of the cross-ratio.

Collecting all the irreducible diagrams we have
\begin{align}
 t_{12}^2t_{34}^2 \partial_{t_1} \partial_{t_2} \partial_{t_3} \partial_{t_4} Q^{(3)}_{irred} = & \
 \frac{8(4 - 8 u + 7 u^2 - 3 u^3 + 2 u^4)}{(1-{\cal B}^2)^2(1-u)^2}+\frac{16 (2 - u + u^4) \log(1 - u)}{(1-{\cal B}^2)^2 u}\nonumber\\
 &
 -\frac{16 u^4 (3 -3u + u^2) \log(u)}{(1-{\cal B}^2)^2 (1- u)^3}
 \nonumber\\
 & -\frac{16t_{12}^2t_{34}^2 }{(1-{\cal B}^2)^3}\left(\frac {3} {t_ {12} t_ {23} t_ {24}^2} - \frac{3}{t_ {12} t_ {13} t_ {14}^2} + \frac{2}{t_ {12} t_ {23}^2 t_ {24}} - \frac{2}{t_ {12} t_ {13}^2 t_ {14}}-\frac{2} {t_ {13} t_ {23} t_{34}^2}
 \right.
 \nonumber\\& \left.
 +\frac{1}{t_{13}t_{23}^2t_{34}}
 +\frac{1}{t_{13}^2t_{23}t_{34}}
 -\frac{1}{t_{12}^2t_{13}t_{14}}
 -\frac{1}{t_{12}^2t_{23}t_{24}}
 \right)\,.
 \end{align}

It is worth noting that the coefficient in front of the anomalous terms becomes $(1-{\cal B}^2)^{-3}$ after collecting all irreducible diagrams. This is in principle a good sign, as other contributions that can potentially cancel these anomalous terms could only come from reducible diagrams with just three propagators. 

\begin{figure}[h!]
    \centering
    \begin{tikzpicture}[scale=0.8]
    \draw[dashed,thick] (0,0) circle [radius=0.8cm];
     \draw[thick] (0.8*0.707,0.8*0.707)
     --(-0.8*0.707,0.8*0.707);
     \draw[thick] (0.8*0.707,0.8*0.707)
     --(0.8*0.707,-0.8*0.707);
     \draw[thick] (-0.8*0.707, 0.8*0.707)
     --(-0.8*0.707,-0.8*0.707);
    \draw[fill=black] (0.8*0.707,0.8*0.707) circle[radius=0.05cm];
   \draw[fill=black] (0.8*0.707,-0.8*0.707) circle[radius=0.05cm];
  \draw[fill=black] (-0.8*0.707,0.8*0.707) circle[radius=0.05cm];
   \draw[fill=black] (-0.8*0.707,-0.8*0.707) circle[radius=0.05cm];
    \node at (0,-1.15) {(a)};
    \end{tikzpicture}
    \hspace{0.4cm}
    %%%%%%%%
    \begin{tikzpicture}[scale=0.8]
    \draw[dashed,thick] (0,0) circle [radius=0.8cm];
     \draw[thick] (0.8*0.707,0.8*0.707)
     --(0.8*0.707,-0.8*0.707);
     \draw[thick] (-0.8*0.707, 0.8*0.707)
     --(-0.8*0.707,-0.8*0.707);
    \draw[thick] (0.8*0.707,-0.8*0.707)
     --(-0.8*0.707,-0.8*0.707);
    \draw[fill=black] (0.8*0.707,-0.8*0.707) circle[radius=0.05cm];
    \draw[fill=black] (0.8*0.707,0.8*0.707) circle[radius=0.05cm];
    \draw[fill=black] (-0.8*0.707,-0.8*0.707) circle[radius=0.05cm];
    \draw[fill=black] (-0.8*0.707,0.8*0.707) circle[radius=0.05cm];
    \node at (0,-1.15) {(b)};
    \end{tikzpicture}
    \hspace{0.4cm}
 \begin{tikzpicture}[scale=0.8]
    \draw[dashed,thick] (0,0) circle [radius=0.8cm];
     \draw[thick] (-0.8*0.707,0.8*0.707)
     --(0.8*0.707,-0.8*0.707);
     \draw[white,fill=white] (0,0) circle [radius=0.1cm];
     \draw[thick] (-0.8*0.707, -0.8*0.707)
     --(0.8*0.707,0.8*0.707);
      \draw[thick] (-0.8*0.707, 0.8*0.707)
     --(0.8*0.707,0.8*0.707); 
    \draw[fill=black] (0.8*0.707,-0.8*0.707) circle[radius=0.05cm];
   \draw[fill=black] (0.8*0.707,0.8*0.707) circle[radius=0.05cm];
    \draw[fill=black] (-0.8*0.707,-0.8*0.707) circle[radius=0.05cm];
  \draw[fill=black] (-0.8*0.707,0.8*0.707) circle[radius=0.05cm];
   \node at (0,-1.15) {(c)};
    \end{tikzpicture}
\hspace{0.4cm}
\begin{tikzpicture}[scale=0.8]
    \draw[dashed,thick] (0,0) circle [radius=0.8cm];
     \draw[thick] (-0.8*0.707,0.8*0.707)
     --(0.8*0.707,-0.8*0.707);
     \draw[white,fill=white] (0,0) circle [radius=0.1cm];
     \draw[thick] (-0.8*0.707, -0.8*0.707)
     --(0.8*0.707,0.8*0.707);
      \draw[thick] (-0.8*0.707, -0.8*0.707)
     --(0.8*0.707,-0.8*0.707);
    \draw[fill=black] (0.8*0.707,-0.8*0.707) circle[radius=0.05cm];
  \draw[fill=black] (0.8*0.707,0.8*0.707) circle[radius=0.05cm];
   \draw[fill=black] (-0.8*0.707,-0.8*0.707) circle[radius=0.05cm];
   \draw[fill=black] (-0.8*0.707,0.8*0.707) circle[radius=0.05cm];
   \node at (0,-1.15) {(d)};
    \end{tikzpicture}
    \hspace{0.4cm}
  \begin{tikzpicture}[scale=0.8]
    \draw[dashed,thick] (0,0) circle [radius=0.8cm];
    \draw[thick,cap=round] (0.8*0.707,0.8*0.707) arc (70:110:1.7cm);
    \draw[thick,cap=round] (0.8*0.707,0.8*0.707) arc (-70:-110:1.7cm);
    \draw[thick] (-0.8*0.707,-0.8*0.707)
     --(0.8*0.707,-0.8*0.707);
    \draw[fill=black] (0.8*0.707,0.8*0.707) circle[radius=0.05cm];
   \draw[fill=black] (0.8*0.707,-0.8*0.707) circle[radius=0.05cm];
   \draw[fill=black] (-0.8*0.707,0.8*0.707) circle[radius=0.05cm];
   \draw[fill=black] (-0.8*0.707,-0.8*0.707) circle[radius=0.05cm];
    \node at (0,-1.15) {(e)};
    \end{tikzpicture}
    \hspace{0.4cm}
 \begin{tikzpicture}[scale=0.8]
    \draw[dashed,thick] (0,0) circle [radius=0.8cm];
    \draw[thick] (-0.8*0.707,0.8*0.707)
     --(0.8*0.707,0.8*0.707);
    \draw[thick,cap=round] (0.8*0.707,-0.8*0.707) arc (70:110:1.7cm);
    \draw[thick,cap=round] (0.8*0.707,-0.8*0.707) arc (-70:-110:1.7cm);
    \draw[fill=black] (0.8*0.707,-0.8*0.707) circle[radius=0.05cm];
    \draw[fill=black] (0.8*0.707,0.8*0.707) circle[radius=0.05cm];
    \draw[fill=black] (-0.8*0.707,-0.8*0.707) circle[radius=0.05cm];
    \draw[fill=black] (-0.8*0.707,0.8*0.707) circle[radius=0.05cm];
    \node at (0,-1.15) {(f)};
    \end{tikzpicture}
\caption{Diagrams contributing to $\partial_{t_1}\partial_{t_2}\partial_{t_3}\partial_{t_4}Q^{(3)}_{red,0}$.}
    \label{fig:enter-label-Q30}
 \end{figure}
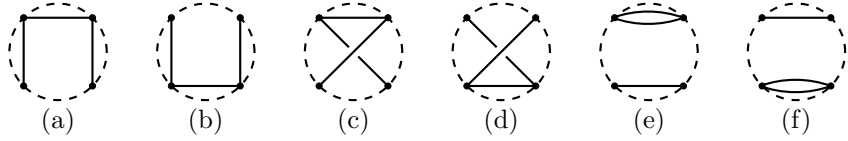

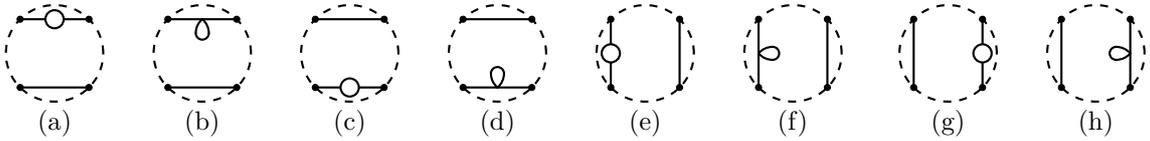
\begin{figure}[h!]
    \centering
    \hspace{-0.2cm}
    %%%%%%%%%%%%%
    \begin{tikzpicture}[scale=0.8]
    \draw[dashed,thick] (0,0) circle [radius=0.8cm];
     \draw[thick] (0.8*0.707,0.8*0.707)
     --(-0.8*0.707,0.8*0.707);
\draw[thick,fill=white] (0,0.8*0.707)
     circle[radius=0.15cm];
    \draw[thick] (-0.8*0.707,-0.8*0.707)
     --(0.8*0.707,-0.8*0.707);
    \draw[fill=black] (0.8*0.707,0.8*0.707) circle[radius=0.05cm];
    \draw[fill=black] (0.8*0.707,-0.8*0.707) circle[radius=0.05cm];
    \draw[fill=black] (-0.8*0.707,0.8*0.707) circle[radius=0.05cm];
    \draw[fill=black] (-0.8*0.707,-0.8*0.707) circle[radius=0.05cm];
    \node at (0,-1.15) {(a)};
    \end{tikzpicture}
    \hspace{0.4cm}
    %%%%%%%%%%%%%
    \begin{tikzpicture}[scale=0.8]
    \draw[dashed,thick] (0,0) circle [radius=0.8cm];
     \draw[thick] (0.8*0.707,0.8*0.707)
     --(-0.8*0.707,0.8*0.707);
         \draw[thick] (-0.8*0.707,-0.8*0.707)
     --(0.8*0.707,-0.8*0.707);
\draw[thick,cap=round] (0,0.8*0.707) arc (50:-0:0.3cm);
\draw[thick,cap=round] (0,0.8*0.707) arc (-230:-180:0.3cm);
\draw[thick,cap=round] (0.108,0.335) arc (0:-180:0.108cm);
    \draw[fill=black] (0.8*0.707,0.8*0.707) circle[radius=0.05cm];
    \draw[fill=black] (0.8*0.707,-0.8*0.707) circle[radius=0.05cm];
    \draw[fill=black] (-0.8*0.707,0.8*0.707) circle[radius=0.05cm];
    \draw[fill=black] (-0.8*0.707,-0.8*0.707) circle[radius=0.05cm];
    \node at (0,-1.15) {(b)};
    \end{tikzpicture}
    \hspace{0.4cm}
     %%%%%%%%%%%%%
    \begin{tikzpicture}[scale=0.8]
    \draw[dashed,thick] (0,0) circle [radius=0.8cm];
     \draw[thick] (0.8*0.707,-0.8*0.707)
     --(-0.8*0.707,-0.8*0.707);
     \draw[thick,fill=white] (0,-0.8*0.707)
     circle[radius=0.15cm];
     \draw[thick] (-0.8*0.707,0.8*0.707)
     --(0.8*0.707,0.8*0.707);
    \draw[fill=black] (0.8*0.707,0.8*0.707) circle[radius=0.05cm];
    \draw[fill=black] (0.8*0.707,-0.8*0.707) circle[radius=0.05cm];
    \draw[fill=black] (-0.8*0.707,0.8*0.707) circle[radius=0.05cm];
    \draw[fill=black] (-0.8*0.707,-0.8*0.707) circle[radius=0.05cm];
    \node at (0,-1.15) {(c)};
    \end{tikzpicture}
    \hspace{0.4cm}
    %%%%%%%%%%%%%
    \begin{tikzpicture}[scale=0.8]
    \draw[dashed,thick] (0,0) circle [radius=0.8cm];
     \draw[thick] (0.8*0.707,-0.8*0.707)
     --(-0.8*0.707,-0.8*0.707);
    \draw[thick,cap=round] (0,-0.8*0.707) arc (-50:-0:0.3cm);
    \draw[thick,cap=round] (0,-0.8*0.707) arc (230:180:0.3cm);
    \draw[thick,cap=round] (0.108,-0.335) arc (0:180:0.108cm);
    \draw[thick] (-0.8*0.707,0.8*0.707)
     --(0.8*0.707,0.8*0.707);
    \draw[fill=black] (0.8*0.707,-0.8*0.707) circle[radius=0.05cm];
    \draw[fill=black] (0.8*0.707,0.8*0.707) circle[radius=0.05cm];
    \draw[fill=black] (-0.8*0.707,-0.8*0.707) circle[radius=0.05cm];
    \draw[fill=black] (-0.8*0.707,0.8*0.707) circle[radius=0.05cm];
    \node at (0,-1.15) {(d)};
    \end{tikzpicture}
     \hspace{0.4cm}
    %%%%%%%%%%%%%
    \begin{tikzpicture}[scale=0.8]
    \draw[dashed,thick] (0,0) circle [radius=0.8cm];
     \draw[thick] (-0.8*0.707, 0.8*0.707)
     --(-0.8*0.707,-0.8*0.707);
    \draw[thick] (0.8*0.707,0.8*0.707)
     --(0.8*0.707,-0.8*0.707);
     \draw[thick,fill=white] (-0.8*0.707,0)
     circle[radius=0.15cm];
    \draw[fill=black] (0.8*0.707,0.8*0.707) circle[radius=0.05cm];
    \draw[fill=black] (0.8*0.707,-0.8*0.707) circle[radius=0.05cm];
    \draw[fill=black] (-0.8*0.707,0.8*0.707) circle[radius=0.05cm];
    \draw[fill=black] (-0.8*0.707,-0.8*0.707) circle[radius=0.05cm];
    \node at (0,-1.15) {(e)};
    \end{tikzpicture}
    \hspace{0.4cm}
    %%%%%%%%%%%%%
    \begin{tikzpicture}[scale=0.8]
    \draw[dashed,thick] (0,0) circle [radius=0.8cm];
     \draw[thick] (0.8*0.707,0.8*0.707)
     --(0.8*0.707,-0.8*0.707);
         \draw[thick] (-0.8*0.707,0.8*0.707)
     --(-0.8*0.707,-0.8*0.707);
\draw[thick,cap=round] (-0.8*0.707,0) arc (140:90:0.3cm);
\draw[thick,cap=round] (-0.8*0.707,0) arc (-140:-90:0.3cm);
\draw[thick,cap=round] (-0.335,0.108) arc (90:-90:0.108cm);
    \draw[fill=black] (0.8*0.707,0.8*0.707) circle[radius=0.05cm];
    \draw[fill=black] (0.8*0.707,-0.8*0.707) circle[radius=0.05cm];
    \draw[fill=black] (-0.8*0.707,0.8*0.707) circle[radius=0.05cm];
    \draw[fill=black] (-0.8*0.707,-0.8*0.707) circle[radius=0.05cm];
    \node at (0,-1.15) {(f)};
    \end{tikzpicture}
    \hspace{0.5cm}
        \begin{tikzpicture}[scale=0.8]
    \draw[dashed,thick] (0,0) circle [radius=0.8cm];
     \draw[thick] (-0.8*0.707,0.8*0.707)
     --(-0.8*0.707,-0.8*0.707);
    \draw[thick] (0.8*0.707,0.8*0.707)
     --(0.8*0.707,-0.8*0.707);
     \draw[thick,fill=white] (0.8*0.707,0)
     circle[radius=0.15cm];
    \draw[fill=black] (0.8*0.707,0.8*0.707) circle[radius=0.05cm];
    \draw[fill=black] (0.8*0.707,-0.8*0.707) circle[radius=0.05cm];
    \draw[fill=black] (-0.8*0.707,0.8*0.707) circle[radius=0.05cm];
    \draw[fill=black] (-0.8*0.707,-0.8*0.707) circle[radius=0.05cm];
    \node at (0,-1.15) {(g)};
    \end{tikzpicture}
    \hspace{0.4cm}
    %%%%%%%%%%%%%
\begin{tikzpicture}[scale=0.8]
    \draw[dashed,thick] (0,0) circle [radius=0.8cm];
     \draw[thick] (-0.8*0.707,0.8*0.707)
     --(-0.8*0.707,-0.8*0.707);
         \draw[thick] (0.8*0.707,0.8*0.707)
     --(0.8*0.707,-0.8*0.707);
\draw[thick,cap=round] (0.8*0.707,0) arc (40:90:0.3cm);
\draw[thick,cap=round] (0.8*0.707,0) arc (-40:-90:0.3cm);
\draw[thick,cap=round] (0.335,0.108) arc (90:270:0.108cm);
    \draw[fill=black] (0.8*0.707,0.8*0.707) circle[radius=0.05cm];
     \draw[fill=black] (0.8*0.707,-0.8*0.707) circle[radius=0.05cm];
    \draw[fill=black] (-0.8*0.707,0.8*0.707) circle[radius=0.05cm];
    \draw[fill=black] (-0.8*0.707,-0.8*0.707) circle[radius=0.05cm];
    \node at (0,-1.15) {(h)};
    \end{tikzpicture}
    \caption{Diagrams contributing to $\partial_{t_1}\partial_{t_2}\partial_{t_3}\partial_{t_4}Q^{(3)}_{red,1}$.}
    \label{fig:enter-label-Q3red1}
\end{figure}

We now turn to reducible diagrams contributing to $Q^{(3)}$. We can distinguish two kinds of reducible contributions:
diagrams with just 3 free propagators and diagrams with 2 propagators, one of them 
being 1-loop corrected. These diagrams are represented in Fig. \ref{fig:enter-label-Q30} and Fig. \ref{fig:enter-label-Q3red1} respectively\footnote{There are more diagrams of these types. In Fig. \ref{fig:enter-label-Q30} and Fig. \ref{fig:enter-label-Q3red1} we show only those non-vanishing after taking the derivatives with respect to the position of the insertion points.}.

Summing diagrams with 3 propagators we find
\begin{align} \nonumber
\partial_{t_1} \partial_{t_2} \partial_{t_3} \partial_{t_4} \, Q^{(3)}_{red,0} & =   \frac{16}{(1 - \mathcal{B}^2)^3} \left [  -\frac{1}{t_{12} t_{23} t_{14}^2}  + \frac{1}{t_{12} t^2_{23} t_{14}} - \frac{1}{t^2_{12} t_{23} t_{14}} 
  + \frac{1}{t_{12} t_{13} t_{24}^2} -\frac{1}{t_{12} t^2_{13} t_{24}}   \right. \\ \nonumber
& -\frac{1}{t_{12}^2 t_{13} t_{24}} +\frac{4}{t^2_{12} t_{34}^2} -\frac{1}{t_{23} t_{14} t_{34}^2} 
-\frac{1}{t_{13} t_{24} t_{34}^2}  
-\frac{1}{t_{23} t_{14}^2 t_{34}} 
+\frac{1}{t^2_{23} t_{14} t_{34}} -\frac{1}{t_{13} t^2_{24} t_{34}} \\
& \left.  
+\frac{1}{t_{13}^2t_{24}t_{34}}  
-\frac12
\left(\frac{1}{t_{23}^2 t_{14}^2} + \frac{1}{t_{13}^2 t_{24}^2} + \frac{2}{t_{12}^2 t_{34}^2} \right)\left(\rm L_{12}+\rm L_{34}\right)
\right]\,.
\end{align}

 For the diagrams in Fig. \ref{fig:enter-label-Q3red1}, with one of propagators corrected at the 1-loop level,
we can use the 2-point function \eqref{2pt1loop} used in section \ref{2point}. We obtain
 \begin{align} \nonumber
\partial_{t_1} \partial_{t_2} \partial_{t_3} \partial_{t_4}  \, Q^{(3)}_{red, 1}   = -\frac{32\pi}{(1 - \mathcal{B}^2)} &\left[ (f_1 - 4f_2)  \left(\frac{1}{t_{23}^2 t_{14}^2} + \frac{1}{t_{13}^2 t_{24}^2} + \frac{2}{t_{12}^2 t_{34}^2} \right)\right. \\
      & \left.+  f_2 \left( \frac{\rm L_{23} + \rm L_{14}}{t_{23}^2 t_{14}^2} + \frac{ \rm L_{13} + \rm L_{24}}{t_{13}^2 t_{24}^2} + \frac{2\rm L_{12} + 2\rm L_{34}}{t_{12}^2 t_{34}^2} \right) \right]\,.
      \label{partialQred1}
\end{align}
We will use, as we have argued in section \ref{2point}, that $f_2 = \frac{1}{4\pi(1-{\cal B}^2)^2}$, but the coefficient $f_1$ will remain indeterminate.

Having computed all the diagrammatic contributions enclosed in $ \partial_{t_1} \partial_{t_2} \partial_{t_3} \partial_{t_4} Q^{(3)}$, it remains to consider the last three terms in \eqref{ddddGS3}, which give
\begin{align} 
d_1 
\partial_{t_1} \partial_{t_2} \partial_{t_3} \partial_{t_4}
\left( G_{\sf S}^{(2)}\left({\rm L}_{12}+{\rm L}_{34}\right)\right) & =
\frac{32}{(1-{\cal B}^2)^3}\left[
-\frac{1}{t_{12} t_{13} t_{14}^2}  
+ \frac{1}{t_{12} t_{23} t_{14}^2}  
- \frac{1}{t_{12} t^2_{13} t_{14}} 
- \frac{1}{t^2_{12} t_{13} t_{14}} 
\right. \nonumber\\
&  
- \frac{1}{t_{12} t_{23}^2 t_{14}}  
+ \frac{1}{t_{12}^2 t_{23} t_{14}}   
-\frac{1}{t_{12} t_{13} t_{24}^2}
+\frac{1}{t_{12} t_{23} t_{24}^2}
+\frac{1}{t_{12} t_{13}^2 t_{24}} 
\nonumber\\
&  
+\frac{1}{t_{12}^2 t_{13} t_{24}}
+\frac{1}{t_{12} t_{23}^2 t_{24}}
-\frac{1}{t_{12}^2 t_{23} t_{24}}
-\frac{1}{t_{13} t_{23} t_{34}^2}
+\frac{1}{t_{23} t_{14} t_{34}^2}
\nonumber\\
&  
+\frac{1}{t_{13} t_{24} t_{34}^2}
-\frac{1}{t_{14} t_{24} t_{34}^2}
+\frac{1}{t_{13} t_{23}^2 t_{34}}
+\frac{1}{t_{13}^2 t_{23} t_{34}}
+\frac{1}{t_{23} t_{14}^2 t_{34}}
\nonumber\\
&  
-\frac{1}{t_{23}^2 t_{14} t_{34}} 
+\frac{1}{t_{13} t_{24}^2 t_{34}}  
-\frac{1}{t_{14} t_{24}^2 t_{34}} 
-\frac{1}{t_{13}^2 t_{24} t_{34}}
-\frac{1}{t_{14}^2 t_{24} t_{34}}
 \nonumber \\
& \left.  
\frac{1}{2}\left(\frac{1}{t_{13}^2t_{24}^2}+\frac{1}{t_{14}^2t_{23}^2}\right)\left({\rm L}_{12}+{\rm L}_{34}\right)\right]\,,
\\
 2 {d_1d_2} \partial_{t_1} \partial_{t_2} \partial_{t_3} \partial_{t_4}\left({\rm L}_{12}{\rm L}_{34}\right) = &
\, \frac{16 d_2}{(1-{\cal B}^2)}
 \frac{1}{t_{12}^2t_{34}^2}\,,
 \label{partialloglog}
\\
\frac{d_1^3}{2}
\partial_{t_1} \partial_{t_2} \partial_{t_3} \partial_{t_4}\left(
{\rm L}_{12}^2{\rm L}_{34}+{\rm L}_{12}{\rm L}_{34}^2\right)= &
 -\frac{128}{(1-{\cal B}^2)^3}
 \frac{1}{t_{12}^2t_{34}^2}
 +\frac{32}{(1-{\cal B}^2)^3}
 \frac{({\rm L}_{12}+{\rm L}_{34})}{t_{12}^2t_{34}^2}\,.
\end{align}

We are in a position to collect all the contributions to $\partial_{t_1} \partial_{t_2} \partial_{t_3} \partial_{t_4}  G^{(3)}_{\sf S}$ and analyze its dependence on $t_1$, $t_2$, $t_3$ and $t_3$. The final result can be simply expressed as follows:
\begin{align}
t_{12}^2t_{34}^2 \partial_{t_1} \partial_{t_2} \partial_{t_3} \partial_{t_4}  G^{(3)}_{\sf S}
   = & P(u)+P\left(\tfrac{u}{u-1}\right)\,,
   \label{finalresult}
\end{align}
where
\begin{align}
    P(u)= &
    -\frac{8 u^2}{(1-{\cal B}^2)^3}
    \left[4 + \log (u^2) \right]
    +\frac{8}{(1-{\cal B}^2)^2}
    \left[ 2+u+2u^2 +
    \left(\tfrac{2}{u} -1 + \tfrac{u^3}{(1-u)^3}\right) \log (u^2) 
    \right]\nonumber\\
    & +\frac{8}{(1-{\cal B}^2)}
    \left[-d_2 +32 \pi f_1\left(1+u^2\right)\right]\,.
\end{align}
As anticipated and in accordance with the conformal symmetry of the line, we observe that the anomalous terms cancel and this is a function of the cross-ratio. Furthermore, crossing symmetry resulting from the exchange between $t_3$ and $t_4$ is evident in \eqref{finalresult} as well.

It is noteworthy to mention the appearance of indeterminate coefficients, such as those in \eqref{partialQred1} and \eqref{partialloglog}. In fact, in the case of $f_2$, the use of the specific value given in section \ref{2point} is crucial for arranging L$_{ij}$ terms  into a function of $u$. Conversely, the terms with $f_1$ and $d_2$ lead to functions of the cross-ratio for any value of these coefficients.

\section{Discussion}
\label{discu}

The type IIA background for the dual description of the ABJ(M) model - AdS$_4\times \mathbb{CP}^3$ - includes a flat Kalb-Ramond field that couples to the string through a boundary term. Therefore, this can affect the boundary conditions of an open string dual to a Wilson loop in the gauge theory. For the case of a straight Wilson line, the dual open string  has an AdS$_2$ world-sheet ending along a line at the boundary of AdS$_4$.  In accordance with the symmetries of 1/6 BPS bosonic Wilson line, the dual open string satisfies certain boundary condition that delocalizes it around a $\mathbb{CP}^1 \subset \mathbb{CP}^3$ \cite{Drukker:2008zx}. One of the main observations of our work is that the fluctuations on the open string dual to the 1/6 BPS bosonic Wilson line in the ABJ(M) model satisfy some boundary conditions mixing longitudinal and transverse derivatives, where the mixing parameter comes from the Kalb-Ramond field. Thus, the Neumann boundary condition - the vanishing of the derivative transverse to the boundary - proposed to describe the smearing of the dual string over a $\mathbb{CP}^1 \subset \mathbb{CP}^3$ \cite{Lewkowycz:2013laa} would be appropriate for the ABJM model, when the two gauge group factors of the Chern-Simons theory has equal ranks. In the ABJ(M) generalization, when the two gauge group factors are different, the open string dual to 1/6 BPS bosonic Wilson loops becomes delocalized around a $\mathbb{CP}^1 \subset \mathbb{CP}^3$ using mixed boundary conditions instead. This constitutes another concrete realization of the supersymmetric mixed boundary conditions (invariant under 4 supercharges) proposed in \cite{CGRS}. 

Using scalar fields with these mixed boundary conditions in the AdS$_2$ world-sheet, we have holographically computed correlation functions of excitations along the 1/6 BPS bosonic Wilson line in the ABJ(M) model. As evidenced by the fact that the quantity \eqref{finalresult} is a function of the cross-ratio, a 1/6 BPS bosonic Wilson line with local operator insertions in the ABJ(M) model  constitutes a CFT$_1$. This result is significant because, being the theory on the line a CFT$_1$, it would be possible to use analytic bootstrap techniques to investigate its correlation functions, extending the analysis conducted in \cite{Bianchi:2020hsz}.  An important difference in the present case is that, despite the supersymmetry of the Wilson loop, some of the fluctuations are not part of protected multiplets and receive corrections to their scaling dimensions. Bootstrap techniques have been shown to be powerful when used in combination with integrability to study Wilson loops in ${\cal N}=4$ super Yang-Mills \cite{Cavaglia:2021bnz,Cavaglia:2022qpg,Cavaglia:2022yvv,Niarchos:2023lot,Cavaglia:2023mmu}. In the present realization of the AdS/CFT correspondence, integrability has been observed for 1/2 BPS Wilson loops in the ABJM setup \cite{Correa:2023lsm}. It would be therefore interesting to explore the integrability for 1/6 BPS Wilson loops in the ABJ(M) model, bearing in mind that the coupling with the Kalb-Ramond field breaks the parity symmetry in the dual line. Some  results of  integrability in ABJ theories for closed  spin chains  can be found for example in \cite{Minahan:2010nn} and references therein.

The mixing boundary conditions we discussed in this article resemble a type of boundary conditions discussed in \cite{Garay:2022szq}. More concretely, the boundary conditions here are similar to the $t$-derivative of the boundary conditions there. However, in that case, as presented in \cite{Garay:2022szq}, the 4-point function does not have the form expected for a conformal field theory. In the current case, a key difference is the fact that the boundary conditions describe a smeared string configuration. This has prompted us to consider orthogonal fluctuations to embedding coordinates. As a result, additional reducible diagram contributions have to be included, which are ultimately the ones cancelling the anomalous terms originated from quartic vertices irreducible diagrams.
 It is important to emphasize that, while the boundary conditions considered in \cite{CGRS} and \cite{Garay:2022szq} were regarded as candidates for the dual string to the interpolating family of Wilson loops \cite{Ouyang:2015iza,Ouyang:2015bmy,Castiglioni:2022yes,Castiglioni:2023uus}, in this current article the boundary conditions we discuss account for the 1/6 BPS bosonic Wilson loops in the ABJ(M) generalization of the model.

Another interesting problem for the future would be the inclusion of Green-Schwarz fermions in the world-sheet, to determine their boundary conditions and to compute fermionic 1-loop Witten diagrams. This would enable a thorough derivation of the 1-loop correction to the propagator in a  detailed fashion. This would be important  to validate the form we have proposed in \eqref{2pt1loop} and ascertain the value for the coefficient $f_2$ in \eqref{expectedf2}, which was crucial for the conformal covariance of both, the 2-point and the 4-point function. Furthermore, one would be able to explicitly determine the coefficients $f_1$ and $d_2$ that appear in the 4-point function and give the anomalous dimension to the next perturbative order. 

\section*{Acknowledgments}
We thank Alberto Faraggi, Mart\'in Lagares and Guillermo Silva for discussion and comments. This work was partially supported by  PICT 2020-03749, PIP 02229 and UNLP X910. DHC would like to acknowledge support from the ICTP through the Associates Programme (2020-2025) during the initial stages of this project.

\appendix

\section{Self-energy corrections to the propagator}
\label{oneloop}

Previously, we have used that the 1-loop correction to the 2-point function takes the form, 
\begin{align}
\hspace{-0.4cm}
 \langle y^A(t_1) y^B(t_2)\rangle_{1,n}=
 {\left(\delta^{AB}- n^A n^B \right)}
 \left(
 f_0+ f_1 {\rm L}_{12} + 
      f_2 {\rm L}_{12}^2\right)+
 {\epsilon^{ABC}n^C}
 {\rm S}_{12}\left(
 g_0  + g_1 {\rm L}_{12}\right)\,.
 \label{2pt1lguess}
\end{align}

\begin{figure}[h!]
    \centering
    \begin{tikzpicture}
    \draw[dashed,thick] (0,0) circle [radius=0.8cm];
    \draw[thick,cap=round]  (-0.8,0) -- (0.8,0);
    \draw[fill=white,thick] (0,0) circle [radius=0.4cm];
    \draw[fill=black] (0.8,0) circle[radius=0.05cm];
    \draw[fill=black] (-0.8,0) circle[radius=0.05cm];
    \node at (0,-1.1) {(a)};
    \end{tikzpicture}
    \hspace{0.95cm}
    \begin{tikzpicture}
    \draw[dashed,thick] (0,0) circle [radius=0.8cm];
    \draw[thick,cap=round]  (-0.8,0) -- (0.8,0);
    \draw[fill=black] (0.8,0) circle[radius=0.05cm];
    \draw[fill=black] (-0.8,0) circle[radius=0.05cm];
    \draw[thick,cap=round] (0,0) arc (-50:-0:0.4cm);
    \draw[thick,cap=round] (0,0) arc (230:180:0.4cm);
    \draw[thick,cap=round] (0.143,0.31) arc (0:180:0.143cm);
    \node at (0,-1.1) {(b)};
    \end{tikzpicture}    
    \caption{Contributions   to $\langle y^A(t_1)y^B(t_2)\rangle_{1,n}$.}
    \label{fig:2pt2app}
\end{figure}
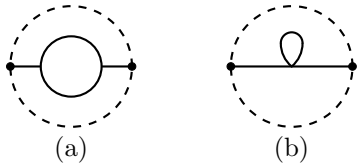

It is possible to study loop diagram contributions to determine whether \eqref{2pt1lguess} is a reasonable assumption or not. There are two types of diagrams contributing to the 1-loop correction of the propagator, as shown in Fig. \ref{fig:2pt2app}. We will briefly analyze the functions obtained from diagrams of the {\it self-energy} type represented by Fig. \ref{fig:2pt2app} (b),  but diagrams \ref{fig:2pt2app} (a), corresponding to {\it fermionic loops} (and harder to compute), will not be considered in this appendix. 
Therefore, this analysis will be focused on verifying the type of functions appearing in \eqref{2pt1lguess} rather than determining the values of the coefficients $f_0$, $f_1$, $f_2$, $g_0$ and $g_1$. 

Let us then concentrate on self-energy diagrams like the one in \ref{fig:2pt2app}(b),
\begin{align}
    \ref{fig:2pt2app}{\rm (b)}_n 
    & =  - \int d^2 \sigma \sqrt{g}  \left \langle
y^A (t_1) \hat{V} (\tau, \sigma) y^B (t_2) \right \rangle
\nonumber\\
   & = \left( \delta^{AB}-n^A n^B\right) \Pi(t_{12}) + \epsilon^{ABC}n^C \Xi(t_{12})\,,
\end{align}
where $\hat{V} (\tau, \sigma)$ represents all the quartic terms in the expansion of the Nambu-Goto-Kalb-Ramond action \eqref{vertices1}-\eqref{vertices2}. The different contributions from each type of vertex depend on which fields are contracted into a tadpole. This will lead to propagators evaluated at coincident points, which have to be regularized to remove UV divergences from them. To this end we can deform ${\rm G}_{\sf s}$ with a covariant UV cutoff
\begin{align}
{\rm G}_{\sf s}(\sigma,\sigma') = & \frac{1}{4\pi}
    \log\left((t-t')^2+ (z-z')^2 + \varepsilon^2 z z'\right)
    +  \frac{1}{4\pi}\frac{1+{\cal B}^2}{1-{\cal B}^2}  \log\left((t-t')^2+ (z+z')^2 \right)\,,
\end{align}
and from this
\begin{align}
 \lim_{\sigma'\to\sigma}
     {\rm G}_{\sf s}(\sigma,\sigma') = & \frac{1}{2\pi }\log(2\varepsilon)+ \frac{1}{\pi} \frac1{1-\mathcal{B}^2}\log z\,,
     \label{nopG} \\
 \lim_{\sigma'\to\sigma}
\partial_z {\rm G}_{\sf s}(\sigma,\sigma') = & \frac{1}{2 \pi z}  \frac{1}{1-\mathcal{B}^2}\,.
\label{pG}
\end{align}
The dependence with the cutoff $\varepsilon$ should cancel once the fermionic loop contribution is included \cite{Beccaria:2019dws}. The double-derivative of ${\rm G}_{\sf s}$ in the coincidence limit is a bit more subtle, as it typically depends on the regularization scheme \cite{Randjbar-Daemi:1987rfs}. A natural choice is to use a scheme that preserves the symmetries of the background \cite{Tseytlin:1988ne}. In the case of Neumann boundary conditions propagator, this choice leads to \cite{Beccaria:2019dws}
\begin{align}
\lim_{\sigma'\to\sigma}
\partial_\mu\partial^{\mu'}{\rm G}_{N}(\sigma,\sigma') = &
\frac{1}{4\pi}
\lim_{\sigma'\to\sigma}
\partial_\mu\partial^{\mu'}
\left[
\log\left((t-t')^2+ (z-z')^2 \right)
+\log\left((t-t')^2+ (z+z')^2 \right)
\right]
\nonumber \\
 = &
 -\frac{1}{2\pi}\,.
 \label{choice}
\end{align}
The second term in this limit is regular and
$\lim_{\sigma'\to\sigma}
\partial_\mu\partial^{\mu'}
\log\left((t-t')^2+ (z+z')^2 \right)
=-1$. This means that the choice made in 
\eqref{choice} implies that
$\lim_{\sigma'\to\sigma}
\partial_\mu\partial^{\mu'}
\log\left((t-t')^2+ (z-z')^2 \right)
=-1 $. This, in turn, leads to
\begin{align}
 \lim_{\sigma'\to\sigma}
\partial_\mu\partial^{\mu'}{\rm G}_{\sf s}(\sigma,\sigma') = & -\frac{1}{2\pi} \frac{1}{1-\mathcal{B}^2} \,.
\label{ppG}
\end{align}

Let us present, with some detail, the following contribution from $V_1$,
\begin{align}
 \label{v1a}
 -\frac{2\pi}{\sqrt{\lambda}}
 \int \frac{d^2\sigma}{z^2}
\langle \wick{
    (\c1 y^C  \c1 y^D   \partial_{\mu} \c1 y^C  \partial^{\mu} \c2 y^D) \c1 y^A (t_1) \c2 y^B (t_2)} \rangle 
& = \frac{2}{\sqrt{\lambda}}
  \frac1{1-\mathcal{B}^2}
    \int \frac{d^2\sigma}{z^2}\log z
    \partial_\mu {\rm K}^{CA} \partial^\mu {\rm K}^{CB}
 \\  
& =
\frac{2}{\sqrt{\lambda}}
  \frac1{(1-\mathcal{B}^2)^3}
  \left[\left( \delta^{AB}-n^A n^B\right) I_1 + \mathcal{B} \epsilon^{ABC}n^C
   I_2\right]  \nonumber 
\end{align}
where we have omitted the $\log\varepsilon$ term. The integrals $I_1$ and $I_2$, up to constants and IR divergences, are\footnote{In the following,
${\rm K}_{N/0}(t_i)$ stands for
${\rm K}_{N/0}(z,t;t_i)$.}
\begin{align}
\label{I1}
I_{1} &= \int \frac{d^2\sigma}{z^2}\log{z}\left(\partial_\mu{\rm K}_N(t_1)\partial^\mu{\rm K}_N(t_2)+{\cal B}^2 \partial_\mu{\rm K}_0(t_1)\partial^\mu{\rm K}_0(t_2)\right) = \frac{(1+{\cal B}^2)}{8\pi} \left(\log16 {\rm L}_{12}-{\rm L}_{12}^2\right)
\\
I_{2} &= \int \frac{d^2\sigma}{z^2}\log{z}\left(\partial_\mu{\rm K}_N(t_1)\partial^\mu{\rm K}_0(t_2) - \partial_\mu{\rm K}_0(t_1)\partial^\mu{\rm K}_N(t_2)\right) = \frac{1}{2} {\rm S}_{12}\left({\rm L}_{12}-\log 4\right)\,.
\end{align}

From this single contribution, we already observe the appearance of terms of the form presented in \eqref{2pt1lguess}. Since we will not consider the fermionic loop contributions here, keeping track of all the self-energy contributions with plenty of details would not be particularly illuminating.  Nonetheless, considering the conformal covariance verification we perform in this article depends on the actual value of the coefficient $f_2$, it might be worthwhile to collect all the contributions of the form ${\rm L}_{12}^2$.

In addition to the one from \eqref{v1a}-\eqref{I1}, other contributions from this vertex to $\Pi$ arise with the coincidence-limit factors \eqref{pG}, \eqref{ppG} and \eqref{ptG}
\begin{align}
\lim_{\sigma'\to\sigma}
\partial_t {\rm G}_{\sf a}(\sigma,\sigma') = & -\frac{1}{2 \pi z}  
\frac{{\cal B}}{1-\mathcal{B}^2}\,.
\label{ptG}
\end{align}
Collecting all of them
\begin{align}
  \Pi_1 = & \frac{1}{\sqrt{\lambda}}
  \left(\frac{2I_1}{(1-{\cal B}^2)^3}+\frac{6I_3}{(1-{\cal B}^2)^3}-\frac{I_4}{(1-{\cal B}^2)^3}-\frac{2{\cal B}^2I_5}{(1-{\cal B}^2)^3}\right)\,,
\end{align}
where now
\begin{align}
I_{3} &=
\int \frac{d^2\sigma}{z}\left({\rm K}_N(t_1)\partial_z{\rm K}_N(t_2)+{\cal B}^2 {\rm K}_0(t_1)\partial_z{\rm K}_0(t_2)\right)
 =\frac{(1+{\cal B}^2)}{8\pi} {\rm L}_{12}^2\,,
 \\
 I_{4} &=
 \int \frac{d^2\sigma}{z^2}\left({\rm K}_N(t_1){\rm K}_N(t_2)+{\cal B}^2 {\rm K}_0(t_1){\rm K}_0(t_2)\right)
 =\frac{(1+{\cal B}^2)}{4\pi}\left({\rm L}_{12}^2+4 {\rm L}_{12}\right)\,,
 \\
 I_{5} &= \int \frac{d^2\sigma}{z}\left({\rm K}_N(t_1)\partial_t{\rm K}_0(t_2)- {\rm K}_0(t_1)\partial_t{\rm K}_N(t_2)\right)
 =  \frac{1}{4\pi} {\rm L}_{12}^2\,.
\end{align}
These contributions come from different Wick contractions of the vertex $V_1$. The $I_3$ and $I_5$ terms comes from 
$\langle \wick{
    (\c1 y^C  \c2 y^D   \partial_{\mu} \c1 y^C  \partial^{\mu} \c1 y^D) \c2 y^A (t_1) \c1 y^B (t_2)} \rangle$ and
    $\langle \wick{
    (\c1 y^C  \c2 y^D   \partial_{\mu} \c2 y^C  \partial^{\mu} \c2 y^D) \c1 y^A (t_1) \c2 y^B (t_2)} \rangle$, while the $I_4$ term comes from 
    $\langle \wick{
    \c1 y^A (t_1) \c2 y^B (t_2)(\c1 y^C  \c2 y^D   \partial_{\mu} \c1 y^C  \partial^{\mu} \c1 y^D) } \rangle$.

Finally, we find 
\begin{align}
    \Pi_1 & = 
    \frac{1}{\sqrt\lambda}
    \frac{1}{8\pi}
    \frac{(1+\mathcal{B}^2)}{(1-\mathcal{B}^2)^3} 
    \left(-2+6-2\right) {\rm L}_{12}^2 -
    \frac{1}{2\pi}
    \frac{\mathcal{B}^2}{(1-\mathcal{B}^2)^3}{\rm L}_{12}^2 +\cdots
    \nonumber\\
    & = \frac{1}{4\pi}
    \frac{1}{(1-\mathcal{B}^2)^2}{\rm L}_{12}^2 +\cdots\,,
\end{align}
where the ellipsis represents the omission of ${\rm L}_{12}$ and constant terms.

We shall not discuss in detail the contributions to $\Pi$ from vertices $V_2$ and $V_3$ with 4 derivatives, nor from vertices $\partial x^i \partial x^i \partial y^C \partial y^C$, as it is straightforward to check that they contribute to the ${\rm L}_{12}$ but not to the ${\rm L}_{12}^2$ dependence.  Other contributions to $\Pi$ could arise from $V_4$, originated from the coupling with the Kalb-Ramond field. For these other contributions, in addition to \eqref{nopG} and \eqref{pG}, we will also need the following propagators in the coincidence limit
\begin{align}
 \lim_{\sigma'\to\sigma}
\partial_{t}\partial_{z'}{\rm G}_{\sf a}(\sigma,\sigma') = & \frac{1}{4 \pi z^2}  \frac{{\cal B}}{1-\mathcal{B}^2}
\,,
\label{ppGa}
    \\
 \lim_{\sigma'\to\sigma}
\partial_t {\rm G}_{\sf a}(\sigma,\sigma') = & -\frac{1}{2 \pi z} \frac{\mathcal{B}}{1-\mathcal{B}^2}\,,
\label{pGa}
\end{align}
The different Wick contractions of $V_4$ give rise to
\begin{align}
  \Pi_4 = & \frac{1}{\sqrt{\lambda}}
  \left(\frac{{\cal B}^2 I_4}{(1-{\cal B}^2)^3}+
 \frac{2{\cal B}^2 I_5}{(1-{\cal B}^2)^3} 
  -\frac{2{\cal B}^2 I_3}{(1-{\cal B}^2)^3}
  +
 \frac{4{\cal B}^2 I_6}{(1-{\cal B}^2)^3}
  \right)\,,
\label{Pi4}
\end{align}
where
\begin{align}
     I_{6} &= \int d^2\sigma\log{z}\left(\partial_z{\rm K}_N(t_1)\partial_t{\rm K}_0(t_2)- \partial_z{\rm K}_0(t_1)\partial_t{\rm K}_N(t_2)\right)
 =  \frac{1}{8\pi} \left(\log16 {\rm L}_{12}-{\rm L}_{12}^2\right)\,.
\end{align}
In \eqref{Pi4}, the terms proportional to $I_4$, $I_5$, $I_3$ and $I_6$ come from
$n^C\epsilon^{CDE}\langle \wick{
    \c1 y^A (t_1) \c2 y^B (t_2) (\c2 y^F  \c1 y^F   \partial_{z} \c1 y^D  \partial_{t} \c1 y^E) } \rangle$,
$n^C\epsilon^{CDE}\langle \wick{
    (\c2 y^F  \c1 y^F   \partial_{z} \c1 y^D  \partial_{t} \c1 y^E)\c1 y^A (t_1) \c2 y^B (t_2)  } \rangle$,
    $n^C\epsilon^{CDE}\langle \wick{
    \c2 y^A (t_1) \c1 y^B (t_2) (\c1 y^F  \c1 y^F   \partial_{z} \c2 y^D  \partial_{t} \c1y^E)  } \rangle$ and
    \\
    $n^C\epsilon^{CDE}\langle \wick{
     (\c1 y^F  \c1 y^F   \partial_{z} \c2 y^D  \partial_{t} \c1y^E) \c2 y^A (t_1) \c1 y^B (t_2)  } \rangle$
respectively. Evaluating \eqref{Pi4} we get
\begin{align}
    \Pi_4 & = 
    \frac{\mathcal{B}^2}{\sqrt\lambda}
    \frac{1}{4\pi}
    \frac{1}{(1-\mathcal{B}^2)^3} 
    \left((1+\mathcal{B}^2)
    +2-(1+\mathcal{B}^2)-2
    \right) {\rm L}_{12}^2 +\cdots
    = 0 +\cdots\,,
\end{align}
Thus, the total contribution to the term proportional to $(\delta^{AB}-n^An^B){\rm L}_{12}^2$ is
\begin{align}
    \Pi =
    \frac{1}{4\pi(1-\mathcal{B}^2)^2}{\rm L}_{12}^2 +\cdots\,.
\end{align}
Therefore, the expected value of the coefficient $f_2$ \eqref{expectedf2} seems to originate from self-energy diagrams exclusively. The same was observed in the case of Neumann boundary conditions \cite{Beccaria:2019dws}.

\end{document}